\documentclass[aps,prl,twocolumn,preprintnumbers,showpacs,amsmath,amssymb,nobalancelastpage]
{revtex4}
\usepackage{amsfonts,mathrsfs,graphicx}
\usepackage{epsfig}
\usepackage{bm}
\usepackage{color}
\def\be{\begin{equation}}
\def\ee{\end{equation}}
\def\bea{\begin{eqnarray}}
\def\eea{\end{eqnarray}}

\begin{document}

\title{Oscillation tomography of the Earth with solar neutrinos 
and future experiments}

\author{
 P.~Bakhti$^{1}$ and  A.Yu.~Smirnov$^{2,3}$
 }
 \address{
 $^1$
School of physics, Institute for Research in Fundamental Sciences (IPM),
P.O.Box 19395-5531, Tehran, Iran\\
$^2$
Max-Planck Institute for Nuclear Physics, Saupfercheckweg 1, 
D-69117 Heidelberg, Germany \\ 
$^3$ ICTP, Strada Costiera 11, 34014 Trieste, Italy \\
}

\begin{abstract}

We study in details the Earth matter effects on the boron neutrinos
from the Sun using recently developed 3D models of the Earth.
The models have a number of new features of the density profiles,
in particular, a substantial deviation from spherical symmetry.
In this connection, we further elaborate on relevant aspects
of oscillations ($\epsilon^2$ corrections, adiabaticity violation, entanglement, {\it etc.})
and the attenuation effect.
The night excesses of the $\nu e-$ and $\nu N-$ events and
the Day-Night asymmetries, $A_{ND}$, are presented in terms of the
matter potential and the generalized energy resolution functions.
The energy dependences of the cross-section and the flux improve the resolution,
and consequently, sensitivity to remote structures of the profiles.
The nadir angle ($\eta$) dependences of $A_{ND}$ are computed for future detectors DUNE,
THEIA, Hyper-Kamiokande, and MICA
at the South pole.
Perspectives of the oscillation tomography of the Earth
with the boron neutrinos are discussed. Next-generation detectors
will establish the integrated day-night asymmetry with high confidence level.
They can give some indications of the $\eta-$
dependence of the effect, but will discriminate among different models
at most at the $(1 - 2)\sigma$ level.
For high-level discrimination, the MICA-scale experiments are needed. MICA can
detect the ice-soil borders and perform unique tomography of Antarctica.

\end{abstract}


\pacs{14.60.Pq, 26.65.+t, 91.35.-x,  95.85.Ry, 96.60.Jw}

\maketitle

\section{I. Introduction}

Oscillations of the solar neutrinos
in the Earth~\cite{before} - \cite{Ioannisian:2017chl} have
the following features.

1. Due to loss of the propagation coherence, the solar neutrinos
arrive at the surface of the Earth as independent fluxes
of the mass eigenstates \cite{ms87},
\cite{Baltz:1988sv,Dighe:1999id, Ioannisian:2004jk}.

2. Inside the Earth, the mass states oscillate in multi-layer
medium with smoothly (adiabatically) changing
density within layers and sharp density change at the borders
between the layers.

3. The oscillations proceed in the low-density regime
which is quantified by a small parameter
\begin{equation}
\epsilon \equiv \frac{2 V E}{\Delta m_{21}^2 },
\end{equation}
where $V(x) = \sqrt{2}G_F n_e(x)$ is the matter potential,
$n_e$ is the electron number density of the medium.
For $E = 10$ MeV at the surface of the Earth
$\epsilon$ equals $\sim 0.03$.

4. The oscillation length
$$
l_m \approx l_\nu \approx 330~ {\rm km} \left( \frac{E}{10 \,{\rm MeV}}\right)
\left(\frac{7.5 \cdot 10^{-5} {\rm eV}^2}{\Delta m^2_{21}}\right)
$$
is comparable to a section of trajectory in a layer $d_i$ for trajectories
with nadir angles $\eta$ close to $\pi/2$:
$d_i = r_i /\cos \eta$, where
$r_i \sim 10$ km is the width of the layer in the
radial direction. The highest sensitivity
is to structures of the density profile of the size $\sim l_m/2$.

5. The attenuation effect is realized in the order $\epsilon$
due to the finite neutrino energy resolution (reconstruction)
in the experimental setup
\cite{Ioannisian:2004vv,Ioannisian:2017chl}. It means loss
of sensitivity to remote structures of the
Earth density profile. Consequently, only structures sufficiently
close to a detector, and therefore to the surface of
the Earth (crust, upper mantle), are most relevant for observations.
This means that with the boron neutrinos,
deep structures, like the core of the Earth, are not seen at the $\epsilon$ level.
The attenuation effect is absent in the order $\epsilon^2$.
Thus, the solar neutrino tomography is essentially
sensitive to the small scale structures in the crust and mantle of the
Earth.

In previous computations, (see, e.g., \cite{Ioannisian:2004jk},
\cite{dnprem})
the density profile of the one-dimensional PREM model
\cite{prem} was used. In this model, borders between layers
have forms of ideal spheres.
Recently several new three dimensional
Earth models have been developed.
They show several new features of the density profiles
which have not been taken into account previously:
(i) the borders between layers are not spherically
symmetric but have irregular deviations from spheres;
(ii) the profiles depend on the azimuthal angle;
(iii) The profiles are non-symmetric with respect
to the center of neutrino trajectory.
The horizontal sizes of these structures are comparable
to oscillation length
which means that effectively they can smooth borders between layers
as well as produce some new parametric effects in oscillations.

In the present paper, we study how these new features
modify the observational effects.
We compute the Earth matter effect using new models.
This allows us to assess the possibility to distinguish the models
with solar neutrino detectors.
At the same time, our computations quantify errors of the computed effects
due to uncertainty in the density profile.

Presently, there is the first (about $3\sigma$) indication of the Earth matter effect
by SuperKamiokande~\cite{Renshaw:2013dzu}, and this situation will stay until
the next generation of experiments will start to operate.
Here we consider solar neutrino studies
by future detectors DUNE \cite{Acciarri:2015uup},
Hyper-Kamiokande (HK)\cite{Hyper-Kamiokande:2016dsw},
THEIA \cite{Alonso:2014fwf, Askins:2019oqj} and MICA \cite{Boser:2013oaa}.

The paper is organized as follows. In Sec.~II, we
present oscillation formalism relevant for our computations and elaborate
on some new features, such as high order $\epsilon$ corrections, entanglement, {\it etc}.
We introduce the generalized energy resolution functions
and study their properties.
The Day-Night asymmetry is presented in terms of these resolution function and potential.
In Sec.~III, new models of the density distribution in the Earth are described.
In Sec.~IV, we present results of computations of the Earth matter effect for
future detectors. Conclusions are given in Sec.~V.

\section{II. \ Relative excess of the night events and attenuation}

\subsection{Coherence and entanglement}

Loss of the propagation coherence is due to spatial separation
of the wave packets that correspond to
the mass eigenstates originated from the same flavor state.
Although separated, these wave packets belong to the same wave function and
therefore entangled. If one of the eigenstates is detected
the parts of the wave function, which describe two other eigenstates, collapse.
It can be easily shown that observational result is the same as
in the case of independent fluxes of mass eigenstates
once total flux of these states is normalized on the total flux
of the originally produced flavor neutrinos.
Coherence is not restored in a realistic detector.

\subsection{Corrections to probability}

Recall that the $\nu_e$ survival probability during a day,
as function of the neutrino energy, equals
\begin{equation}
\label{PD}
P_D (E) = \frac{1}{2} {c_{13}^4}\left[1 +
\cos 2 \theta_{12} \cos 2 \bar{\theta}^m_{12}(E) \right] + s_{13}^4,
\end{equation}
where $c_{13} \equiv \cos \theta_{13}$, $s_{13} \equiv \sin \theta_{13}$,
and $\cos 2\bar{\theta}^{m}_{12}$
is the mixing parameter averaged over the boron neutrino
production region in the Sun \cite{Bahcall:1996qv}:
\begin{equation}
\cos 2\bar{\theta}^m_{12} \ \approx \
{\cos 2 \theta_{12} - c_{13}^2 \bar{\epsilon}_\odot
\over
\sqrt{(\cos 2 \theta_{12} - c_{13}^2 \bar{\epsilon}_\odot)^2 +
\sin^2 2 \theta_{12}}}.
\label{cosi}
\end{equation}
Here
\be
\label{epsi}
\bar{\epsilon}_\odot \equiv \frac{2 \bar{V}_\odot E}{\Delta m_{21}^2 }
\ee
and $\bar{V}_\odot$ is the averaged matter potential in the $^8$B neutrino production region.

For high energy part of the
boron neutrino spectrum, where $\bar{\epsilon}_\odot \gg 1$,
we have
\begin{equation}
\cos 2 \theta_{12}^m (E) \approx - \left[1 -
\frac{\sin^2 2 \theta_{12}}{2c_{13}^4}
\left(\frac{\Delta m^2_{21}}{2 \bar{V}_\odot E}\right) \right].
\end{equation}
So, dependence on $E$ is weak. At the solar neutrino energies
the matter effect on the 1-3 mixing is negligible,
therefore $ \bar\theta_{13} \approx \theta_{13}=8.4^\circ$
\cite{13mix}.

During a night the probability equals $P_N = P_D + \Delta P$, where
the difference of the night and day probabilities
is given to the order $\epsilon^2$ by
\cite{Ioannisian:2004vv,Ioannisian:2017dkx}
\begin{eqnarray}
\Delta {P} (E) =
\kappa(E) \left[\int_{0}^{L} \! dx
\ V(x) \sin \phi^m ( L - x, E) + I_2 \right].
\label{eq:diffprob}
\end{eqnarray}
Here
$$
\kappa (E) \equiv -\frac{1}{2}c_{13}^6 \cos 2 \bar{\theta}^\odot_{12}(E)
\sin^2 2 \theta_{12} \approx 0.5
$$
is slowly changing function of $E$, and
\begin{equation}
I_2 \equiv \frac{1}{2}\cos 2\theta_{12}\left [\int_{0}^{L} \! dx
\ V(x) \cos \phi^m (L - x) \right]^2
\label{eq:secondod}
\end{equation}
is a correction of the order $\epsilon^2$, since
in (\ref{eq:secondod}) each integral over $x$
is of the order $\epsilon$.
The integration in (\ref{eq:secondod}) proceeds along a neutrino trajectory.
In new models of the Earth apart from the nadir angle $\eta$
the density and potential profiles depend,
also on position of the detector ${\bf x}_D$
and azimuthal angle $\phi_a$: $V = V(x, {\bf x}_D, \eta, \phi_a)$.
Correspondingly, for a given detector and
a given of moment of time $\Delta P = \Delta P ({\bf x}_D, \eta, \phi_a)$.

In Eq. (\ref{eq:diffprob})
\begin{equation}
\phi^m (L-x, E) \equiv \int_{x}^{L} \! \! d x \ \Delta_{21}^m(x).
\label{phasexl}
\end{equation}
is the adiabatic phase acquired from a given point of trajectory $x$ to a detector at $L$.
$\Delta_{21}^m(x)$ is
the level splitting and in our calculations we use
it up to the first order in $\epsilon$:
\begin{eqnarray}
\Delta_{21}^m = \Delta_{21} \sqrt{(\cos 2\theta_{21} -
c_{13}^2\epsilon)^2+\sin^22\theta_{21}}
\nonumber\\
\approx \Delta_{21} (1- c_{13}^2 \cos 2 \theta_{12} \epsilon).
\nonumber
\end{eqnarray}
Here $\Delta_{21} \equiv \Delta m_{21}^2 /2E$ is the splitting in vacuum.
Consequently, the oscillation phase (\ref{phasexl}) equals
\begin{equation}
\phi^m (L-x, E) = \Delta_{21} \left[ (L-x) -
c_{13}^2 \cos2 \theta_{12} \int_x^L dx \epsilon (x) \right].
\label{phasexl1}
\end{equation}
Introducing the average density along a neutrino trajectory
$\bar{\rho} (\eta)$, we can rewrite Eq.~(\ref{phasexl1}) as
\begin{equation}
\phi^m (L-x, E) =
\phi^m_0 + \delta \phi^m
\label{phasexl2}
\end{equation}
where $\phi^m_0 \equiv \Delta_{21} (L-x) $ is the zero order phase and
\begin{equation}
\delta \phi^m = \phi^m_0 c_{13}^2 \cos2 \theta_{12} \epsilon (\bar{\rho}),
\label{deltaf}
\end{equation}
is the phase shift due to the $\epsilon-$ correction.

For $\Delta m^2=7.5 \times10^{-5}$~eV$^2$ and
$\rho = 5$ g/cm$^3$ the relative size of the correction
(second term in Eq. (\ref{phasexl2})) is about $3\%$.
For large $\phi^m_0$ the phase shift $\delta \phi^m$ can be observable.
E.g., if $\phi^m_0 = 5 \pi$, we find $\delta \phi^m = 27^{\circ}$.

The correction $\delta \phi^m$ leads
to the shift of oscillatory pattern
in the $\eta$ scale. Since $\delta \phi^m = \Delta_{21} \delta L(\eta)$
and $L = 2R \cos \eta$ we obtain
\begin{equation}
\delta \eta = \frac{\delta \phi^m}{2R \sin \eta \Delta_{21}}.
\label{eq:deleta}
\end{equation}
Insertion of expression for $\delta \phi^m$ (\ref{deltaf}) in to
(\ref{eq:deleta}) gives
\begin{equation}
\delta \eta = \cot \eta ~ c_{13}^2 \cos 2\theta_{12} \epsilon.
\end{equation}
For $\eta = 70^{\circ}$ we obtain $\delta \eta = 0.2^{\circ}$, while period
of oscillatory dependence in the $\eta$ scale for this
$\eta$ equals $2.8^{\circ}$, {\it i.e.}, the shift is by 1/14 of the period.
$\delta \eta$ increases with decrease of $\eta$.

Let us consider $I_2$ -- the second term in (\ref{eq:diffprob}).
For constant density it can be computed explicitly
\begin{equation}
I_2 \approx 0.5 \cos2\theta_{12} \bar{\epsilon}^2
\sin^2 (L\Delta_{21}).
\label{eq:attenuation3}
\end{equation}
Apart from $\bar{\epsilon}^2$, this term contains
additional small factor $0.5 \cos 2\theta_{12} \approx 1/6$.
As a result, $I_2$ is about $0.015\%$ and therefore can be neglected.
Our computational relative errors are of the order of $0.1\%$.
Thus, the largest correction to the probability follows from $\phi^m$.

\subsection{Comments on adiabaticity}

In the lowest order in $\epsilon$,
the sensitivity to structures of the Earth matter profile,
its deviation from constant density,
appears due to borders between layers which strongly (maximally)
break adiabaticity. Indeed, in the adiabatic case
the oscillation probability would depend on density at the surface
of the Earth and on the oscillation phase.
However, in the lowest (zero) order in $\epsilon$
the phase coincides with the vacuum phase.
The matter correction to the phase
is proportional to $\epsilon$ which then appears
as $\epsilon^2$ in the probability.
So, in the adiabatic case, there is no sensitivity to the profile
in the $\epsilon$ order.

In general, deviations of borders between layers from spherical form may produce effective
smearing of borders for neutrino trajectories with large $\eta$, and consequently, to
decrease of the adiabaticity violation. That would lead to
partial loss of sensitivity to the density profile.

If deviation from spherical form in radial direction, $\Delta h$,
and in horizontal direction, $l_{f}$, are such that neutrino
trajectory at certain $\eta$ crosses the border between the same layers several (many) times,
the density gradient along the trajectory will decrease.
For density jump in a border $\Delta \rho$ the gradient equals $\Delta \rho \cos \eta/ \Delta h$.
The scale of density change
$$
l_\rho \equiv \rho (d\rho /dl)^{-1} =
\frac{\rho}{\cos \eta \Delta \rho} \Delta h
$$
should be compared with the oscillation length
in the adiabaticity condition.

As we will see, typical scale of deviation of, e.g., the border between the crust and mantle
from spherical form is $ \Delta h \sim 5 - 10$ km and the horizontal size of the
structures is $l_f \sim (70 - 150)$ km. This gives the slope
of the structure $\eta_f \sim \Delta h/l_f \sim (2- 7)^\circ$. Therefore
double crossing can occur for the trajectories with
$\eta > 83^\circ$.
For parameters of new Earth models,
however, adiabaticity is still strongly broken and multiple crossing
of borders can occur only in very narrow intervals of $\eta$.

In the lowest $\epsilon$ order, the result for $\Delta P(E)$ in (\ref{eq:diffprob})
can be reproduced as a result of interference of the ``oscillation waves" emitted from
borders between layers \cite{Ioannisian:2017dkx}.
For $i$th wave, the phase is determined by distance from border to a detector
$L - x_i$ and vacuum oscillation length, while the amplitude is proportional by the density
jump $\Delta \rho_i$ in the border. Then $\Delta P(E)$ is the sum of the waves over
borders which neutrino trajectory crosses. This representation gives simple interpretation
of results of numerical computations.

\subsection{Attenuation and generalized energy resolution functions}

The Earth matter effect can be quantified by the Day-Night asymmetry
or the relative excess of night events (events rate) in energy range
$\Delta E$ as function of the nadir angle $\eta$:
\begin{equation}
A_{ND}(\eta, \Delta E) \equiv
\frac{\Delta N_N (\eta, \Delta E)}{N_D(\Delta E)}, \, \, \,
\Delta N_N \equiv N_N - N_D.
\label{eq:d-nasym}
\end{equation}
Here $N_N (\eta)$ and $N_D$ are the numbers of night and day
events (rates) correspondingly.
The nadir
$\eta$ and azimuthal $\phi_a$ angles are fixed by the detection
time of an event. According to new models, $N_N (\eta)$ depends also on the
position of a detector.

In experiments, the observables are the electron energy and direction.
Therefore, $\Delta E $ is determined by the observed energy interval
of the produced (or recoil) electrons. In practice, we will use the
energy of electrons above certain threshold.
Thus, information on the density profile is
encoded in the nadir angle
dependence of the night excess. We will not consider
the direction of electron.

Sensitivity of oscillations to the Earth density profile
is determined by the sensitivity of a given experimental
set-up to the true energy of neutrino $E$. This can be described by
the generalized energy resolution function $G_\nu (E^r, E)$
such that
\begin{equation}
\Delta N (E^r) = D \int dE~ G_\nu (E^r, E) \Delta P(E),
\label{eq:ndiff}
\end{equation}
where $E^r$ is the observed (reconstructed)
neutrino energy or certain energy characteristic which
can be measured in experiment.
In (\ref{eq:ndiff}) $D$ is the factor which includes characteristics of detection:
fiducial volume, exposure time, {\it etc.} It cancels in the expression
for the relative excess $A_{ND}$.
The resolution function is normalized as
$\int G_\nu(E^r, E) dE = 1$.
Similarly, one can write expression for $N_D$.

$G_\nu(E^r, E)$ includes
the neutrino energy resolution function: $g_\nu (E^r, E)$,
the energy dependence of the neutrino
flux $f_B (E)$~\cite{Bahcall:1996qv} and cross-section $\sigma (E)$:
\begin{equation}
G_\nu (E^r, E) \propto g_\nu(E^r, E) \sigma (E) f_B(E).
\end{equation}
It should also include the energy dependent efficiency of detection.

Integration over the neutrino energy with the resolution
function in Eq. (\ref{eq:ndiff}) leads to the attenuation
effect \cite{Ioannisian:2004jk,Ioannisian:2017chl}.
Plugging expression for $\Delta P(E)$ from (\ref{eq:diffprob})
into (\ref{eq:ndiff}) and neglecting $I_2$ we obtain for $\Delta N$
\begin{equation}
D \int_0^L dx V(x) \int_0^{E^{max}} dE~ G_\nu (E^r, E)
\sin \phi^m ( L - x, E).
\label{eq:permut}
\end{equation}
Here integrations over $x$ and $E$ are interchanged.
In this form the dependence of difference of events
on structures of density profile is immediate.

Let us introduce the attenuation factor $F(L-x)$ \cite{Ioannisian:2004jk}
such that the integral over $E$ in Eq. (\ref{eq:permut}) equals
\begin{eqnarray}
& & \hskip-1cm \int dE G_\nu(E^r, E) \sin \phi^m (L - x, E)
\nonumber \\
& & = F(L-x) \sin \phi^m (L - x, E^r).
\label{eq:attenf}
\end{eqnarray}
In general, this equality can not be satisfied,
but it is valid for special cases and under integral over $x$.
Then the expression for $\Delta N$ in (\ref{eq:permut}) becomes
\begin{equation}
\Delta N(E^r) = D \int dx V(x) F(L-x) \sin \phi^m (L - x, E^r).
\label{eq:permut1}
\end{equation}

For the Gaussian form of $G_\nu(E^r, E)$,
the attenuation factor is given by
\begin{equation}
F(d)\simeq e^{-2\left({ d \over \lambda_{att} }\right)^2},
\label{eq:attfactorx}
\end{equation}
where
\begin{equation}
\lambda_{att} \equiv l_\nu \frac{E}{\pi \sigma_E}
\label{eq:attlength}
\end{equation}
is the {\it attenuation length}, and $l_\nu$ is the oscillation
length in vacuum
\begin{equation}
l_\nu = \frac{4\pi E}{\Delta m_{21}^2}.
\label{eq:osclength}
\end{equation}
According to (\ref{eq:permut1}) and (\ref{eq:attfactorx})
for $d \gg \lambda_{att}$ the attenuation factor $F(d) \approx 0$,
and therefore contributions of remote structures to the integral
(\ref{eq:permut1}) and therefore to observable
oscillation effect is suppressed.
For $d = \lambda_{att}$ the factor $F(d) = e^{-2} \approx 0.14$,
and the attenuation becomes significant.
Consequently, the Day-night asymmetry depends mainly on the shallow
structures of the Earth which are close to a detector.

For the ideal resolution, $G_\nu (E^r, E) = \delta (E^r - E)$,
Eq.~(\ref{eq:attenf}) gives
$F(L-x) = 1$, which means that attenuation is absent.

The attenuation length is the distance at which oscillations
integrated over the energy resolution interval
$\sigma_E$ are averaged out, or the difference of the oscillation phases for
$E$ and $E + \sigma_E$ becomes larger than $2\pi$ \cite{Ioannisian:2017chl}.

Expression (\ref{eq:permut}) factorizes different dependences:
The generalized resolution function encodes external characteristics:
neutrino flux, cross-section, energy resolution of a detector.
$V(x)$ gives information about the density profile, oscillation
probability is reduced to $\sin \phi^m$.

In what follows we will find
expressions for the generalized reconstruction functions
and present numbers of events in the form (\ref{eq:permut})
separately for the $\nu-$nucleon and $\nu-e$ scattering.

\subsection{Neutrino-nuclei scattering}

We consider the charged current neutrino-nuclei interactions and the corresponding resolution function
$G_{\nu N}$.
If transitions to excited states are neglected, the energies of electron
and neutrino are uniquely related (upto negligible nuclei recoil):
$E_e = E - \Delta E$. Here $\Delta E \approx \Delta M + m_e$
is the threshold of reaction.
If transitions to excited states are significant but the energy of de-excitation is not measured, an additional
uncertainty in reconstruction of the neutrino energy appears which should be included into
$G_{\nu N}$.

The night-day difference of numbers of events
with the observed energy of electron $E^r_e$ is given by
\begin{equation}
\Delta N (E^r_e) = D \int_0^{E_e^{max}} \hskip-0.5cm dE_e g_e (E_e^r, E_e)
\sigma (E) f_B (E) \Delta P(E),
\label{eq:nddiff}
\end{equation}
where $E = E_e + \Delta E$, $E_e^{max}$ is maximal true energy of electron:
$E_e^{max} = E^{max} - \Delta E$,
$g_e (E_e^r, E_e)$ is the electron energy resolution
function with $E_e$ and $E_e^r$ being
the true and the observed energies correspondingly.

Introducing also $E^r \equiv E_e^r + \Delta E$
and changing integration in (\ref{eq:nddiff})
to integration over
the neutrino energy $E$ we have
\begin{equation}
\Delta N (E^r_e) = D \int_{\Delta E}^{E^{max}} dE g_\nu (E^r, E)
\sigma (E) f_B (E) \Delta P(E),
\label{eq:nddiff1}
\end{equation}
where $g_\nu (E^r, E) \equiv g_e (E^r - \Delta E, E - \Delta E)$.
The equation (\ref{eq:nddiff1}) can be rewritten as
\begin{equation}
\Delta N (E^r_e) = D z \sigma (E^r) f_B (E^r)
\int_0^{E^{max}} \hskip-0.5cm dE~ G_{\nu N}(E^r, E) \Delta P(E),
\label{eq:nddiff2}
\end{equation}
with
\begin{equation}
G_{\nu N} (E^r, E) = z^{-1} g_\nu (E^r, E)
\frac{ \sigma (E) f_B (E)}{ \sigma (E^r) f_B (E^r)},
\label{eq:ggen}
\end{equation}
and $z$ being the normalization factor.
Inserting expression for $\Delta P(E)$ from (\ref{eq:diffprob}) into
(\ref{eq:nddiff2}) and permuting integrations over $x$ and $E$ we obtain
\begin{eqnarray}
\Delta N (E^r_e) & = & D z \sigma (E^r) f_B (E^r) \kappa(E^r) \times
\nonumber\\
& & \hskip-2cm \int dx V(x) \int_0^{E_\nu^{max}} \hskip-0.3cm dE
~G_{\nu N} (E^r, E) \sin \phi^m(x, E),
\label{eq:nddiff2a}
\end{eqnarray}

Integration over the energy can be removed introducing
of the attenuation factor, as in
(\ref{eq:attenf}), which gives
\begin{eqnarray}
\Delta N (E^r_e) & = & D z \sigma (E^r) f_B (E^r) \kappa(E^r) \times
\nonumber\\
& & \hskip-0.5cm \int dx V(x) F_{\nu N} (L - x) \sin \phi^m(x, E^r).
\label{eq:nddiff2b}
\end{eqnarray}
Finally, integration over the interval of observed energies of electrons gives
\begin{eqnarray}
\Delta N (\Delta E^r_e) & = & D z
\int_{E^{th}}^{E^{max}} dE^r \sigma (E^r) f_B (E^r) \kappa(E^r)
\nonumber\\
& & \hskip-1cm \times \int dx V(x) F_{\nu N} (L - x) \sin \phi^m (x, E^r),
\label{eq:nddigg}
\end{eqnarray}
where we again substituted integration over $E_e$ by integration over $E$.

For the day signal, which does not depend practically on $\eta$,
we have
\begin{eqnarray}
N_D (\Delta E^r_e) & = & D z
\int_{E^{min}}^{E^{max}} \hskip-0.3cm dE^r \sigma (E^r) f_B(E^r) P_D(E^r)
\nonumber\\
& & \times \int_{\Delta E}^{E^{max}} dE G_{\nu N} (E^r, E).
\label{eq:nddiff2c}
\end{eqnarray}
Notice that if threshold $\Delta E$ is low enough, the second integral
over the resolution function
is $\approx 1$, so that
\begin{equation}
N_D (\Delta E^r_e) = D z
\int_{E^{min}}^{E^{max}} \hskip-0.2cm dE^r \sigma (E^r) f_B(E^r) P_D(E^r).
\label{eq:nddiff3}
\end{equation}
The factors $D z$ cancel in the expression for $A_{ND}$.

\begin{figure}[h]
\begin{center}
\includegraphics[width=0.45\textwidth,height=0.35\textwidth]{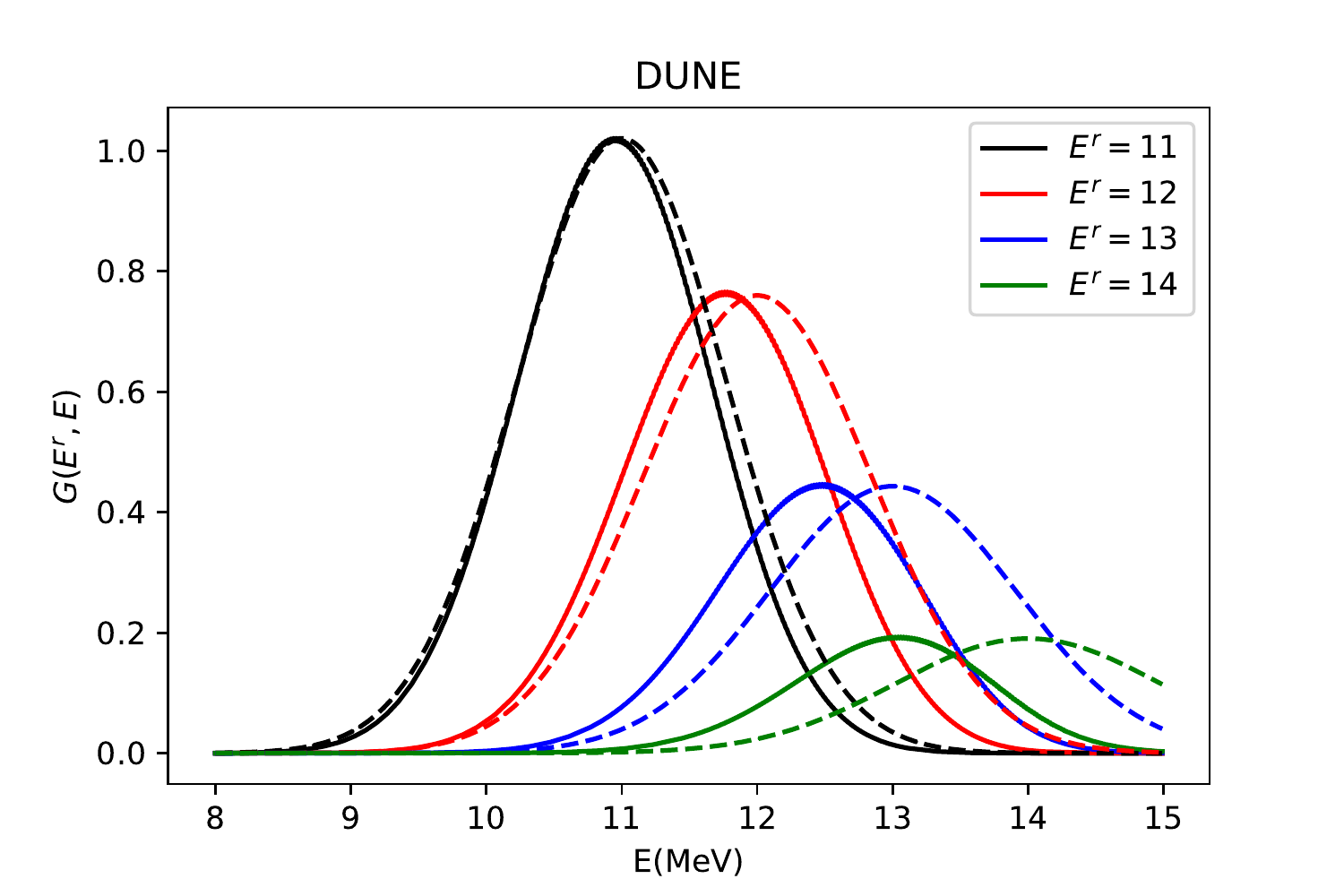}
\includegraphics[width=0.45\textwidth, height=0.35\textwidth]{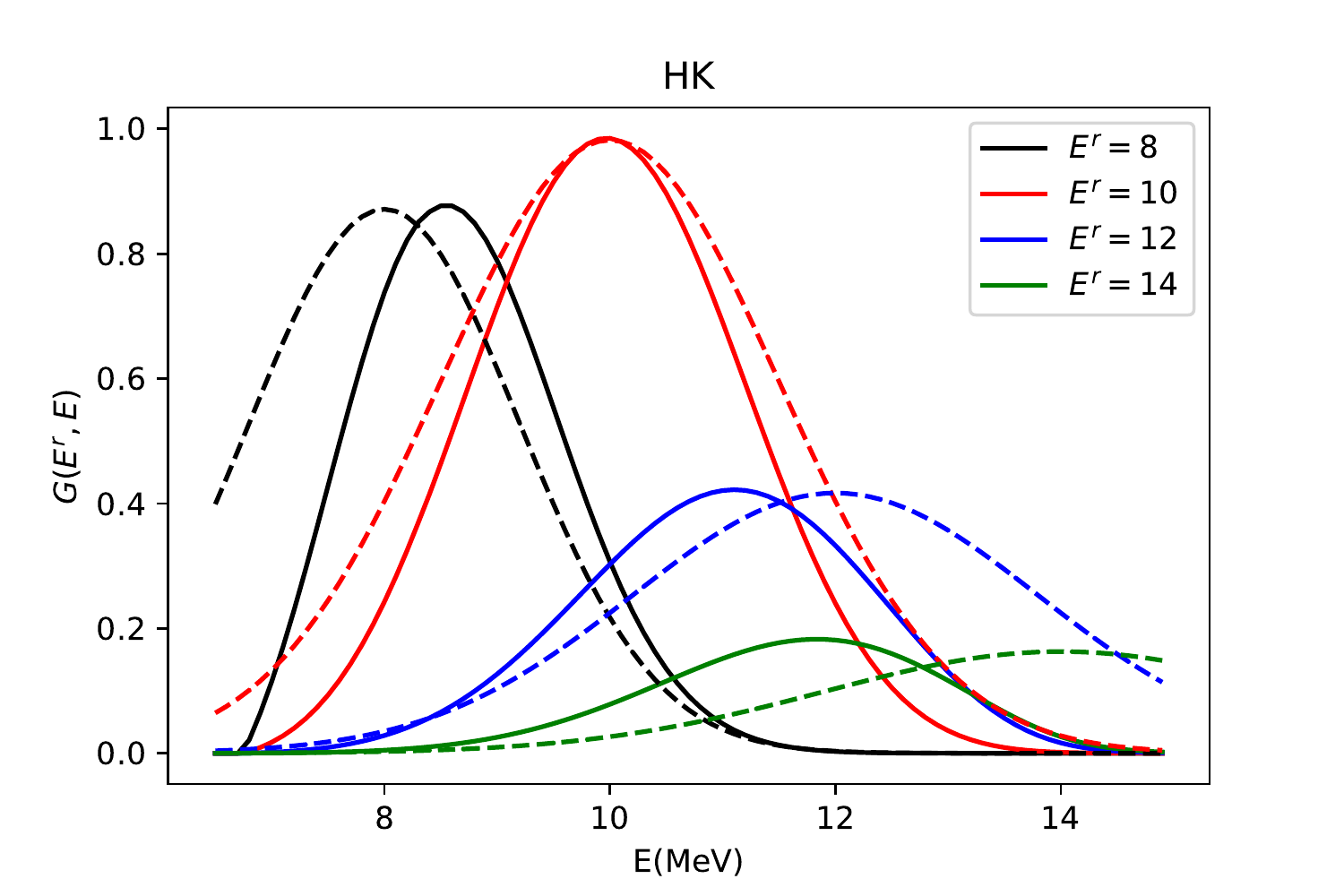}
\end{center}
\caption[...]{
Dependences of the generalized neutrino energy resolution functions
$G_\nu (E^r, E)= g_\nu (E^r, E)\sigma(E)f_B(E)$ (solid lines) and
$G_\nu (E^r, E) = g^{Gauss} (E^r, E)$ (dashed lines)
on true neutrino energy
$E$ for different values of the reconstructed neutrino energy $E^r$.
For $g_\nu(E^r, E)$ we take the Gaussian distribution with width
$\sigma_E$.
{\it The upper panel:} $G_{\nu N}$ - for experiments based on
the $\nu_e - $ nuclei scattering with $\sigma_E = 0.07 E^r$ (DUNE).
{\it The bottom panel:} $G_{\nu e}$ - for experiments based on
the $\nu - e$ scattering with $\sigma_E = 0.15 E^r$ (HK) and the cut
$E_e > 6.5$~MeV.
\label{gauss}
}
\end{figure}

Let us consider the generalized energy resolution function
$G_{\nu N} (E^r, E)$ in details.
In the expression for $G_{\nu N} (E^r, E)$ in Eq. (\ref{eq:ggen}), we use
(i) $\sigma \propto E p$, (ii)
the Gaussian function for $g_\nu (E^r, E)$
with central energy $E^c = E^r$ and the energy resolution
$\sigma_E = 0.07 E_r$ (as for DUNE), (iii)
the flux of Boron neutrinos, $f_B(E)$
from \cite{Bahcall:1996qv}.
Fig.~\ref{gauss} (upper panel) shows by solid lines
dependence of $G_{\nu N}$ on energy $E$
for several values of $E^r$.
We compare this dependence with Gaussian form
$g^{Gauss} (E^r, E)$ (dashed lines) computed with the same $E^r$ and $\sigma_E$.
For convenience of comparison, we normalized $G_{\nu N}(E^r, E)$
in such a way that $G_{\nu N}(E^r, E)^{max} =
g^{Gauss}(E^r, E)^{max}$; and the y-axis is in arbitrary unit.


The figure illustrate effect of inclusion of energy
dependence of $\sigma(E)$ and $f_B(E)$
into resolution function.
The product $\sigma(E) f_B(E)$ has the form
of a wide asymmetric peak with maximum at $\sim 11$ MeV.
Consequently, for $E^r = 11$ MeV the generalized function $G_{\nu N}$
is close to the corresponding Gaussian form with energy of maximum
$E^c \approx E^r$, while
for $E^r > 11$ MeV the factor $\sigma(E) f_B(E)$ shifts $G_{\nu N}$
to lower energies, $E^c < E^r$, and reduces the width. According to Fig.~\ref{gauss}
for $E^r = 12$ MeV the energy of maximum $E^c = 11.7$ MeV and the relative
width $\sigma_E/ E = 0.06$ instead of 0.07 in $g_\nu^{Gauss}$.
The change becomes more profound with increase of $E^r$.
For $E^r = 14$ MeV we find $E^c = 13.1$ MeV and $\sigma_E/ E = 0.05$.
Thus, the energy dependence of $\sigma f_B$ leads to better energy resolution
and therefore to increase the attenuation length which means
the improvement of sensitivity to remote structures.


Notice that inclusion of $\sigma f_B $ into $G_{\nu N}$, not only gives a shift
of the peak and decrease of width, but also changes the shape of the resolution
function which becomes asymmetric.
Still, according to Fig.~\ref{gauss}, for Gaussian $g_\nu$,
the whole resolution function $G_{\nu N}$
can be well approximated by the Gaussian
function with appropriately chosen energy of maximum,
$E^c = E^c(E_r) \neq E_r $, and width $\sigma_E = \sigma (E^r)$.
{\it A priory}, the form of $g_\nu (E^r, E)$ is not known, and
eventually will be determined in experiment.
Therefore in our computations we will use the
generalized reconstruction function in the Gaussian form:
\begin{equation}
G_{\nu N} (E^r, E) \approx
g^{Gauss} [E, E^{max}(E^r), \sigma(E^r)].
\label{eq:gpr-app}
\end{equation}

Under integration over the neutrino energy $E$ the difference of results
for $A_{ND}$ computed with the Gaussian $G_{\nu N}$
(\ref{eq:gpr-app}) and
$G_{\nu N} $ with Gaussian $g_\nu $ is negligible.
Using the PREM model we find that
the relative difference results for $A_{ND}$
is smaller than 0.3$\%$.

\subsection{Neutrino-electron scattering}

In this case the energies of neutrino and electron are not uniquely related,
but correlated via the differential cross-section $d\sigma(E, E_e)/dE_e$.
Correspondingly, expression for the effective resolution function in (\ref{eq:permut})
will differ from $G_{\nu N}$. 

The difference of numbers of the night and day events
with a given observed energy of electron $E_e^r$ equals
\begin{eqnarray}
\Delta N(E_e^r) & = & D \int_{0}^{E^{max}} dE_e g_e (E_e^r, E_e)
\nonumber\\
& & \hskip-1cm \times \int_{E_e}^{E^{max}} dE \frac{d \sigma_\Delta (E, E_e)}{d E_e}
\Delta P(E) f_B(E),
\label{eq:nend}
\end{eqnarray}
where
\begin{equation}
\frac{d \sigma_\Delta (E, E_e)}{d E_e} \equiv
\frac{d\sigma_e (E, E_e)}{d E_e} - \frac{d\sigma_{\mu} (E, E_e)}{dE_e}
\label{eq:diffxsec}
\end{equation}
is the difference of the $\nu_e e$, $d\sigma_e/dE_e$,
and $\nu_\mu e$, $d\sigma_\mu/dE_e$, differential cross-sections.
Interchanging integrations over $E_e$ and $E$ in
Eq. (\ref{eq:nend}) we obtain
\begin{equation}
\Delta N (E_e^r) =
D \int_{0}^{E^{max}} \hskip-0.5cm dE \Delta P(E) f_B(E) \sigma_\Delta(E) g_\nu (E_e^r, E),
\label{eq:nend2}
\end{equation}
where
\begin{equation}
g_\nu (E_e^r, E) \equiv \frac{1}{\sigma_\Delta(E)} \int_{0}^{E} dE_e
\frac{d \sigma_\Delta (E, E_e)}{d E_e} g_e (E_e^r, E_e),
\label{eq:gnu1}
\end{equation}
and
\begin{equation}
\sigma_\Delta (E) = \int_{0}^{E} dE_e
\frac{d \sigma_\Delta (E, E_e)}{d E_e}.
\label{eq:cross2}
\end{equation}
The generalized reconstruction function
can be introduced similarly to (\ref{eq:ggen}):
\begin{equation}
G_{\nu e}(E_e^r, E) = z^{-1} g_\nu (E, E_e^r)
\frac {f_B(E) \sigma_\Delta(E)}
{f_B(E_e^r) \sigma_\Delta(E_e^r)},
\label{eq:genrece}
\end{equation}
or explicitly, inserting $g_\nu$ from (\ref{eq:gnu1}), as
\begin{eqnarray}
G_{\nu e}(E_e^r, E) & = &
\frac{z^{-1} f_B(E)}{f_B(E_e^r)\sigma_\Delta(E_e^r)}
\nonumber\\
& & \hskip-0.5cm \times \int_{0}^{E} dE_e
\frac{d \sigma_\Delta (E, E_e)}{d E_e} g_e (E_e, E_e^r).
\label{eq:genrece1}
\end{eqnarray}
The only difference from (\ref{eq:ggen})
is that here in $g_\nu$ the electron resolution function is integrated
with the differential cross-section.

Instead of $E_e^r$ we can introduce the ``observable"
neutrino energy $E^r = E^r(E_e^r)$ defined as the
energy of maximum of $G_{\nu e}$ for a given $E_e^r$:
\begin{equation}
G_{\nu e}(E_e^r, E^r) = G_{\nu e}^{max} (E_e^r).
\label{eq:obsnuen}
\end{equation}
In terms of $G_{\nu e}(E^r_e, E)$ the N-D difference of numbers
of events can be presented as
\begin{eqnarray}
\Delta N (E_e^r) & = & D z f_B (E^r(E_e^r)) \sigma_\Delta(E^r(E_e^r)) \times
\nonumber\\
& & \int_{0}^{E^{max}} dE \Delta P(E) G_{\nu e}(E^r(E_e^r), E).
\label{eq:nend3}
\end{eqnarray}
As in the $\nu N-$ case,
we insert explicit expression for $\Delta P(E)$ and
interchange integration over $x$ and $E$.
Then the integration over $E$ can be removed introducing
the attenuation factor which gives
\begin{eqnarray}
\Delta N (E_e^r) & = &
D z f_B (E^r) \sigma_\Delta(E^r) \kappa(E^r) \times
\nonumber\\
& & \hskip-0.5cm \int dx V(x) F_{\nu e} (L - x) \sin \phi^m (x, E^r),
\label{eq:nddiffen}
\end{eqnarray}
where $F_{\nu e}(L - x)$ corresponds to $G_{\nu e} (E^r, E)$.

The difference of numbers of events with the observable
energy of electrons in the interval
$\Delta E_e^r \equiv (E_e^{r, min} - E_e^{r, max})$ equals
\begin{eqnarray}
\Delta N (\Delta E_e) & = & D z
\int_{E_e^{r, min}}^{E^{r, max}} dE_e^r f_B (E^r)
\sigma_\Delta(E^r) \kappa(E^r)
\nonumber\\
& & \hskip-0.5cm \int dx V(x) F_{\nu e} (L - x) \sin \phi^m (x, E^r),
\label{eq:neint}
\end{eqnarray}
and $E^r = E^r(E_e^r)$ is determined by (\ref{eq:obsnuen}).

The number (rate) of events with the observed electron
energy $E_e^r$ during a day equals
\begin{eqnarray}
N_{D}(E_e^r) & = &\int_{0}^{E^{max}} dE f_B(E)
\left[P_D (E) \sigma^e (E, E_e^{th}) g_\nu^e(E_e^r, E) \right.
\nonumber\\
& + & \left. (1 - P_D(E)) \sigma^\mu (E, E_e^{th})
g_\nu^\mu(E_e^r, E) \right].
\label{eq:nev3}
\end{eqnarray}
Here
\begin{equation}
g_\nu^{e, \mu} (E_e^r, E) \equiv
\int_{0}^{E} dE_e \frac{d \sigma_{e, \mu} (E, E_e)}{\sigma_e(E) d E_e}
g_e(E_e^r, E_e).
\label{eq:nev2e}
\end{equation}
The total cross-sections are given by
$$
\sigma_{e, \mu} (E) = \int_{0}^{E} dE_e
\frac{d \sigma_{e, \mu} (E, E_e)}{d E_e}.
$$

Expression (\ref{eq:nev3}) can be simplified
assuming $g_\nu^\mu \approx g_\nu^e \approx g_\nu$:
\begin{eqnarray}
N_{D}(E_e^r) & = &\int_{0}^{E^{max}} \hskip-0.5cm dE f_B(E) g_\nu (E_e^r, E)
\left[P_D (E) \sigma^e (E, E_e^{th}) \right.
\nonumber\\
& + & \left. (1 - P_D(E)) \sigma^\mu (E, E_e^{th}) \right].
\label{eq:nev3sim}
\end{eqnarray}

Let us consider $G_{\nu e}(E, E_e^r)$ in detail.
In the bottom panel of Fig.~\ref{gauss}
we show $G_{\nu e}(E, E_e^r)$ as function of $E$ computed according to
Eq.~(\ref{eq:genrece1}). 
We take the Gaussian form for $g_\nu (E^r, E)$
with central energy $E^c = E^r$ and the energy resolution
$\sigma_E = 0.15 E_r$. 
For the $\nu - e$ scattering the product $\sigma(E) f_B(E)$
has wide peak with maximum at $E~=~10$ MeV, and additional
weak $E$ - dependence comes from the integral in (\ref{eq:genrece1}).
Therefore the smallest deviation of $G_{\nu e}(E^r, E)$ from
the Gaussian form is at $E^r \sim 10$ MeV.
For $E^r < 10$ MeV the maximum of $G_{\nu e}$ is shifted to higher energies, while
for $E^r > 10$ MeV -- to lower energies. In both cases the width of $G_{\nu e}$ decreases.
According to Fig.~\ref{gauss} (bottom)
for $E^r = 8$ MeV the maximum of $G_{\nu e}$
is shifted with respect to $E^r$ to higher energy
by $0.5$ MeV, and the width is slightly smaller.
For $E^r = 12$ MeV, inversely, the maximum is shifted
to $E^c = 11.3$ MeV, and the width becomes $\sigma_E / E = 0.12$.
This trend (due to fast decrease of the flux with energy
above 10 - 11 MeV) is even more significant for larger $E^r$:
at $E^r = 14$ MeV, we find $E^c = 11.9$ MeV and $\sigma_E / E = 0.11$.
Again, taking into account the energy dependence of $\sigma$ and $f_B$
improves the energy resolution, but this improvement is weaker than in the $\nu N$ case.

The biggest contribution to oscillation effect comes from the energy range (10 - 12) MeV,
where $G_{\nu e}$ is rather close to the Gaussian form.
Therefore in computations, we will
use the Gaussian form for $G_{\nu e}$ with modified $E^c$ and $\sigma_E$,
and consequently, the attenuation factor in the form (\ref{eq:attfactorx}).
Inclusion of the flux and cross-section energy dependences
narrows the resolution function.

In expressions for
$\Delta N$ the $\phi_a$ dependence appears in
two places: in the potential: $V = V(x, \eta, \phi_a)$ and in the
phase $\phi^m = \phi^m (\phi_a)$.
For each $\eta$ and position of the detector we performed
averaging of $\Delta N$
over the azimuthal angle $\phi_a$.
If $\phi_a$ dependence of the phase is neglected,
in the first approximation, the averaging of $\Delta N$ over $\phi_a$
is reduced to averaging of the potential.

\section{III. \ Models of the Earth and Density profiles}

In computations, we used density profiles
reconstructed from recently developed 3D models of the Earth.
Due to the attenuation effect, the Day-Night asymmetry
mainly depends on shallow density structures:
crust, upper mantle and crust-mantle border called Moho,
or Mohorovicic discontinuity.
There are two types of crust: the oceanic crust and
the continental one. The width of oceanic crust is about (5 - 10) km,
while the continental crust is thicker: (20 - 90) km \cite{moho,Moho1}.
The predicted depth of Moho, $h_{Moho}$, significantly
varies for different models. In contrast, the density change in the Moho
is nearly the same for all the models.
Beneath Homestake the jump is from 2.9 gr/cm$^3$ to 3.3 gr/cm$^3$.

A brief description of relevant elements of the models
is given below.

\begin{figure}[h]
\begin{center}
\includegraphics[width=0.45\textwidth, height=0.35\textwidth]{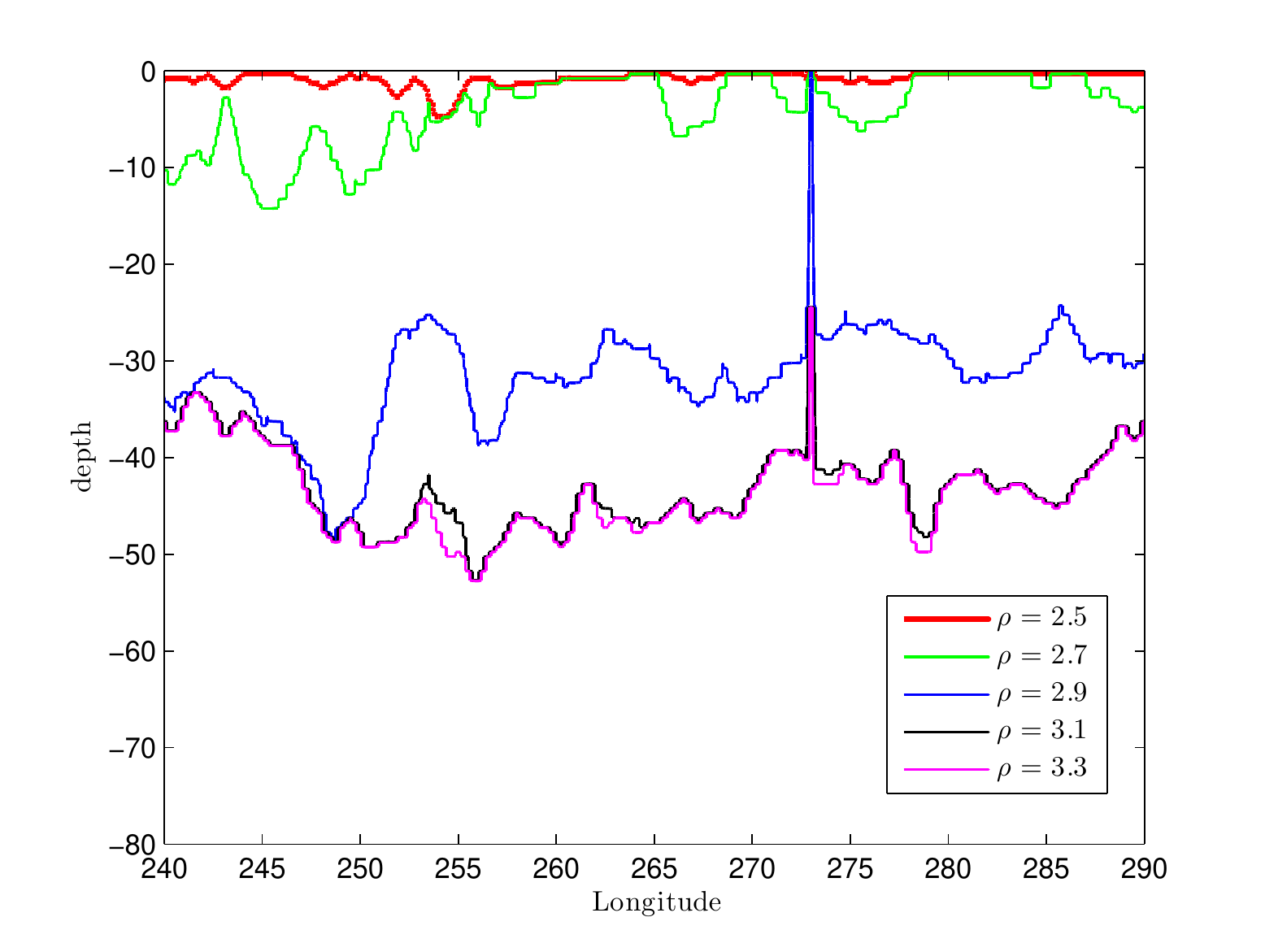}\\
\includegraphics[width=0.45\textwidth, height=0.35\textwidth]{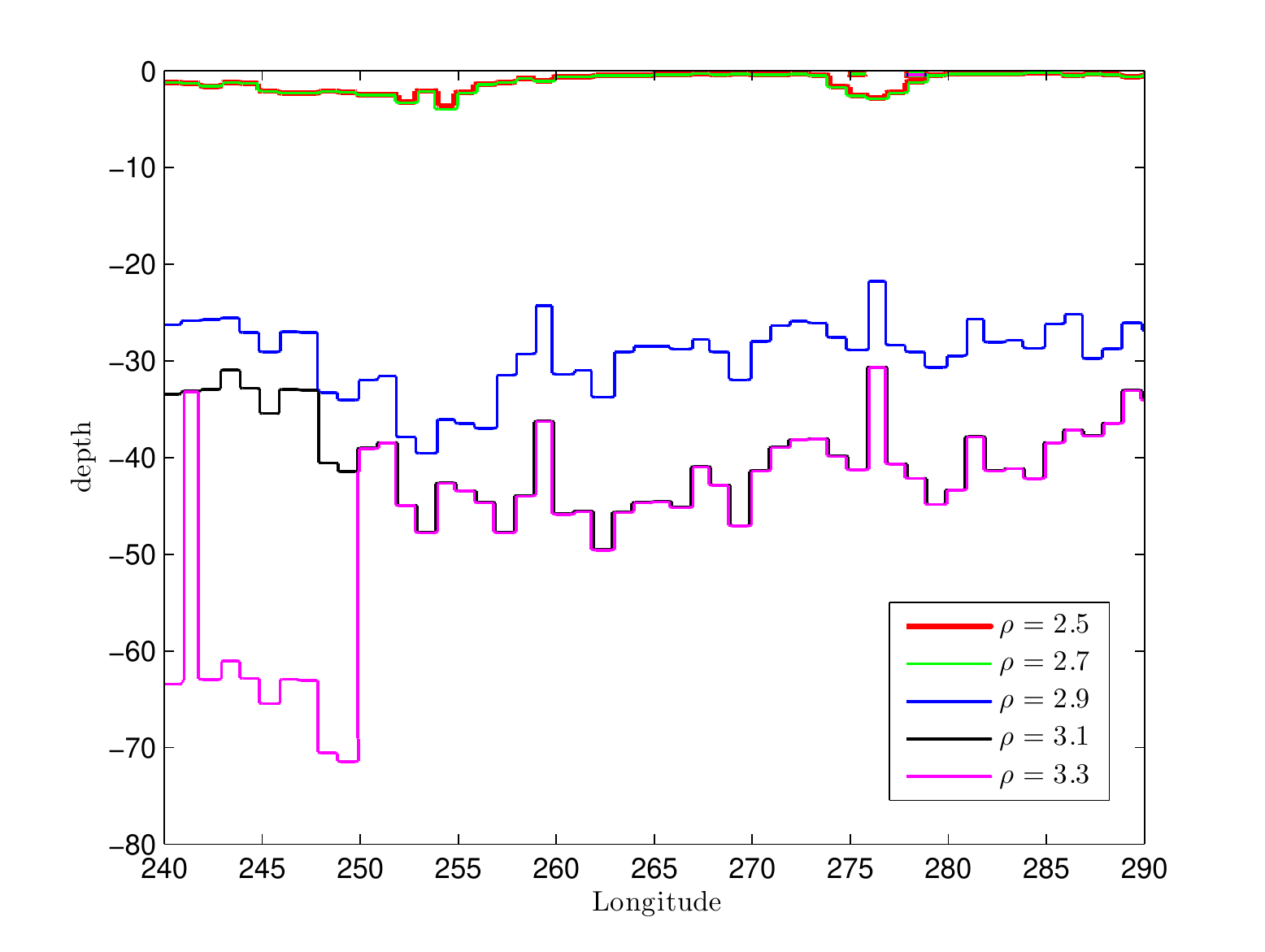}\\
\end{center}
\caption[...]{
The depth of layers with several fixed densities
beneath the Homestake mine (the latitude $44.35^\circ$)
as function of the azimuthal angle (longitude).
The upper panel: Shen-Ritzwoller model, the  bottom panel: the CRUST1 model.
The black lines show the depth of Moho.
\label{densityus}
}
\end{figure}

1. The Shen-Ritzwoller model (S-R) \cite{Shen}
is based on joint Bayesian Monte Carlo inversion
of geophysical data. It gives the density profile
of the crust and uppermost mantle
beneath the US, in area with latitudes $(20^\circ - 50^\circ)$
and longitudes $(235^\circ - 295^\circ)$. In the radial direction
it provides the density change from the sea level surface down
to the depth of 150~km with $h_{Moho} = 52$ km
beneath the Homestake (see Fig. \ref{densityus}).

2. FWEA18, the Full Waveform Inversion of East Asia model \cite{FWEA18},
covers the latitudes $10^\circ - 60^\circ$ and
longitudes $90^\circ - 150^\circ$.
It gives the density profile from the surface down to 800~km,
and $h_{Moho} = 33$ km beneath Kamioka.

3. SAW642AN \cite{SAW642AN} is a global (all latitudes and
longitudes) radially anisotropic mantle shear velocity model
based on a global three-dimensional tomography of the Earth.
The model gives the density profile of mantle starting
from the depth of Moho, $h_{Moho} = 24$ km, down to 2900~km.
No crust structure is available.

4. CRUST1 \cite{crust1} is a global 3D model,
that presents data with 1$\times$1 degree grid in latitude
and longitude at the surface.
It gives the density and depth of borders of eight layers
of the crust: water, ice, upper sediments,
middle sediments, lower sediments, upper crust, middle crust,
lower crust. The model predicts the depths of Moho $h_{Moho} = 48$ km
and $h_{Moho} = 40$ km beneath Homestake and Kamioka respectively and
nearly constant density of the upper mantle down to 100 km.
It provides also the density distribution above the sea level.

Using these models we reconstructed the density, and consequently $V(x)$,
profiles along neutrino trajectories determined by
position of detectors, $\eta$ and $\phi_a$.
Maximal depths $h^{max}$ down to which the models provide data
are $h^{max}({\rm S-R}) = 150$ km, $h^{max} ({\rm CRUST1}) \approx 80$ km,
$h^{max}({\rm FWEA18}) = 800$ km, $h^{max}({\rm SAW642AN}) = 2900$ km.
Therefore we reconstructed the density profiles using the following
prescription:

\begin{itemize}

\item
for the S-R, CRUST1 and FWEA18 models
with relatively small $h^{max}$
we take the SAW642AN profile in the range $h = h^{max} - 2900$ km.

\item
Below 2900 km for all the models we use
the PREM profile. Recall that PREM - the Preliminary
reference Earth model is a one-dimensional model that
represents the average (over solid angle) density of the Earth
as a function of depth. The depth of Moho in the PREM model
equals $h_{Moho} = 24.4$ km.

Due to attenuation effect possible uncertainties related to these compilations
of the profiles do not change results significantly even for small nadir angles.

\item
For purely mantle model SAW642AN above Moho, $h = (0 - 24)$ km,
we take constant density $\rho = \rho_{SAW} (24 ~{\rm km})$.

\end{itemize}

All the models, but CRUST1, give the density below the sea level.
In all simulations, except the case of MICA,
we consider the surface of Earth as perfect sphere and
take zero density above the sea level.
Effect of these simplifications is much smaller than sensitivity
of all experiments (but MICA) due to restricted statistics.
In the case of MICA, we have taken into account the Earth structures
above sea level.

In Fig.~\ref{densityus}, we present the S-R and CRUST1 density profiles
beneath Homestake for fixed latitude $44.35^\circ$.
Both models provide data for this place down to 80 km.
Shown is the depth of layers with a given density as
function of longitude (azimuthal angle).
Notice that at the latitude $44^\circ$ the $1^\circ$ of longitude
corresponds to 76 km at the surface. The black curves show Moho depth, where
density jumps approximately from 2.9 to 3.3 g/cm$^3$.

Few comments are in order.

1. The surfaces of equal density, and in particular, borders between
layers deviate from spherical form.

2. There are irregular deviations from spherical form with
typical angular size $(2 - 5)^\circ$ or $(150 - 400)$ km,
which is comparable with the oscillation length.
The depth variation, $\delta h$, is up to (5 - 10) km,
{\it i.e.} up to $30\%$.

3. There are narrow spikes of large amplitude and wide
regions $\sim 10^\circ$, where the depth increases by $30 \%$
with respect to average value.

4. Two models give rather similar density distributions:
the average depths and lengths
are similar. At the same time,
variations of S-R and CRUST1 models are not correlated.

In the case of spherical inner structures the nadir angle $\eta_c$
at which neutrino starts to cross a given border between
layers with the depth $h$ equals
\begin{equation}
\sin \eta_c = 1 - \frac{h}{r_E},
\label{eq:eatc}
\end{equation}
where $r_E = 6371$ km is the radius of the Earth.
For $\eta < \eta_c$ neutrino crosses this border twice.
Neutrino ``sees'' the mantle for the first time at
$\eta_{Moho} = 83.6^\circ$ in the S-R model,
at $\eta_{Moho} = 83.4^\circ$
in the CRUST1 model and at $\eta_{Moho} = 84.9^\circ$
in the SAW642AN model on September 23
(where the date fixes the azimuthal angle).

The noticeable difference between the S-R (CRUST1) profile
and SAW642AN profile
appears above the S-R Moho depth $h > 52$ km.
Below S-R Moho all three models give similar results.

According to Fig. \ref{densityus}
there are deviations of Moho
from of ideal sphere of two types:

1) Relatively small variations of $2 - 5^\circ$
scale which would correspond to
(150 - 400) km at the DUNE latitude and the size (depth) $\pm (2 - 5)$ km.

2) Long (continental) scale variations of size $50^\circ$
with depth 20 km such that the smallest depth,
$h_{min} = 32$ km, is close to ocean and the bigger depth
$h_{max} = 52$ km is in the center of continent.
This means that the Moho border varies within the shell
(we call it Moho shell)
restricted by spherical surfaces with depth $32 - 52$ km and average
depth 42 km.

The length of neutrino trajectory within the Moho shell equals
$\approx 2\sqrt{2 r_E (h_{max} - h_{min}) } \approx 710$ km which is 2 times
bigger than the oscillation length.
According to (\ref{eq:eatc}) borders of the Moho shell are seen from a detector site
at $\eta_{min} = 84.2^\circ$ and $\eta_{max} = 82.7^\circ$. So that
for $\eta > \eta_{min}$ there is no crossings of Moho: in the interval
$\eta = (\eta_{min} - \eta_{max}) $ one may expect multiple crossing of Moho and
since horizontal scale of variations of the border is comparable to the oscillation length,
parametric effects are expected.
However, averaging over azimuthal angle washes out these effects.
For $\eta < \eta_{max}$ neutrino trajectory crosses the Moho shell twice,
and within each crossing, it can be more than one crossing of the Moho border.
Substantial effect due to Moho crossings is expected at
$\eta \sim 83^{\circ}$.

Below 83$^\circ$ neutrinos cross the Moho in all the models.
For smaller $\eta$
the differences in these models become small.

\begin{figure}[h]
\begin{center}
\includegraphics[width=0.45\textwidth, height=0.35\textwidth]{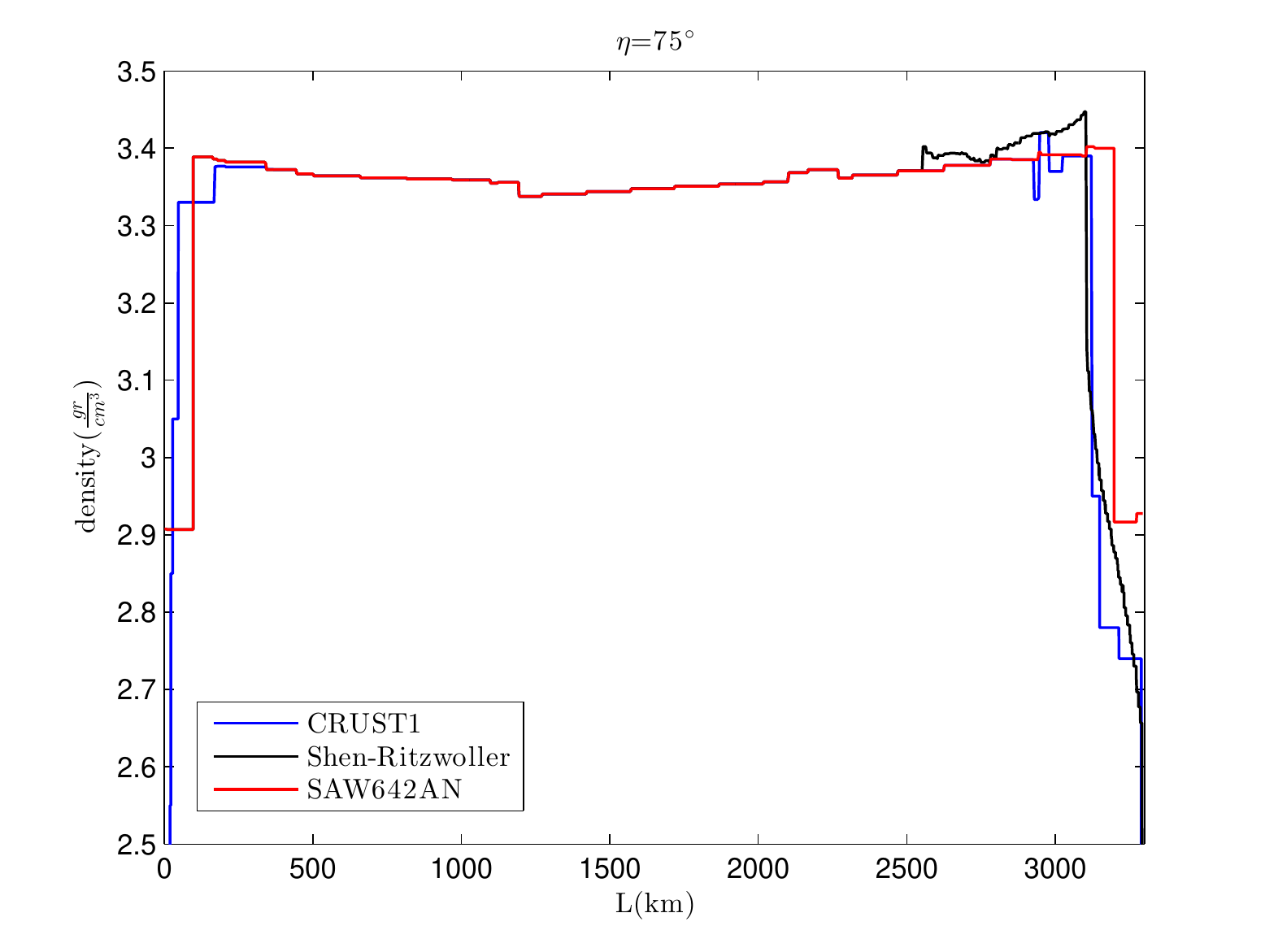}\\
\end{center}
\caption[...]{
The density of the Earth along the neutrino trajectory at nadir angle 75$^\circ$,
and detector in Homestake mine as a function of distance
from the point of entering the Earth.
\label{density-ha}
}
\end{figure}

As an example, in Fig.~\ref{density-ha},
we show the reconstructed density profiles
of three models along the neutrino
trajectory which ends at Homestake with $\eta=75^\circ$
on September 23. The length of trajectory equals 3295~km.
According to Fig.~\ref{density-ha} neutrinos cross the Moho border
second time after 3055~km at a depth of 46~km in the S-R model.
For CRUST1 model the corresponding numbers are
3121~km and 43~km, while for SAW642AN model they equal
3198~km and 24~km.

\begin{figure}[h]
\begin{center}
\includegraphics[width=0.45\textwidth, height=0.35\textwidth]{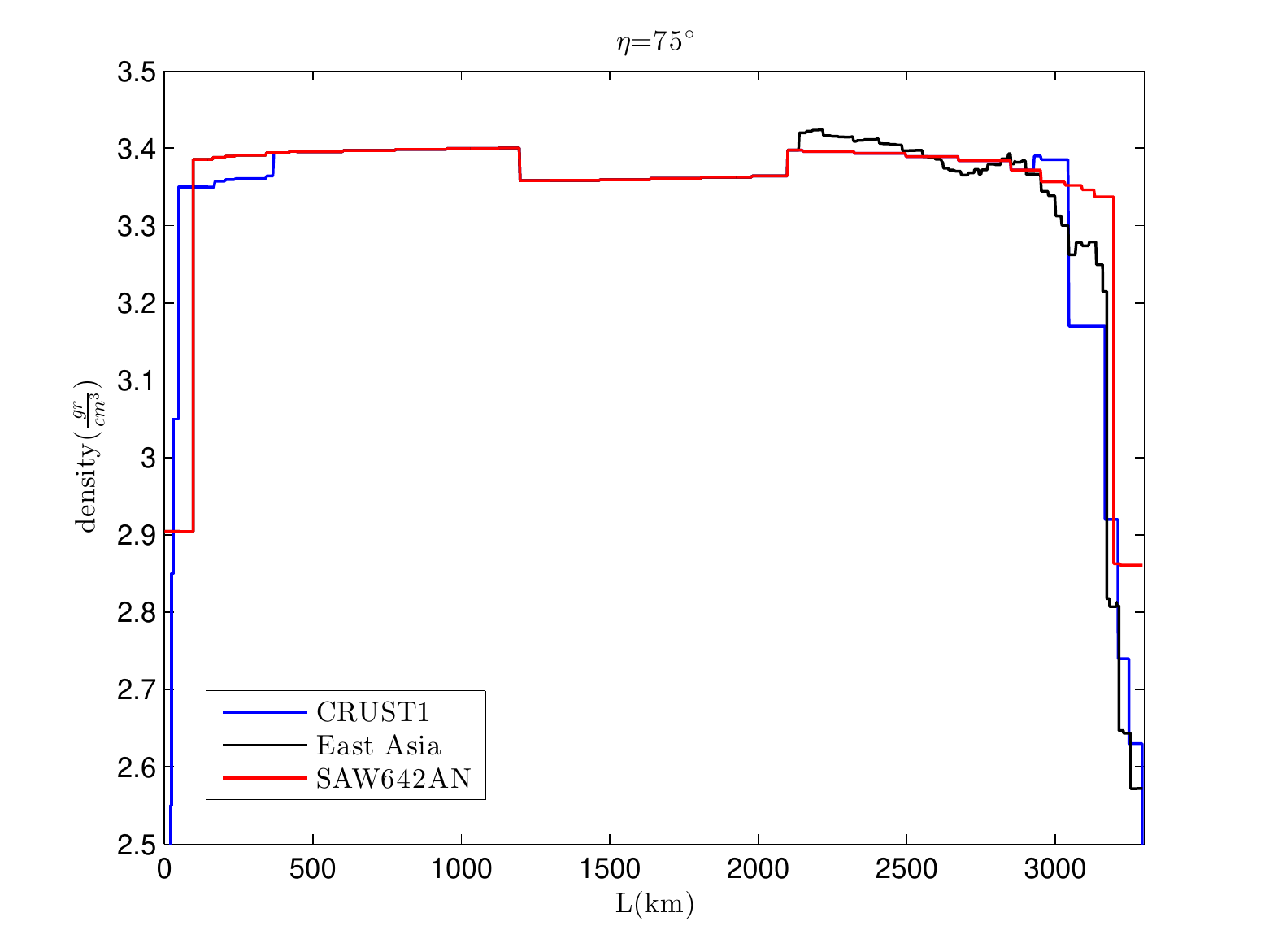}\\
\end{center}
\caption[...]{
The same as in Fig. \ref{density-ha} but for
the detector located at Hida.
\label{densityhk}
}
\end{figure}

In Fig.~\ref{densityhk},
similar profiles are shown at the Hida place and or nadir angle $75^{\circ}$.

Clearly, the profiles are not symmetric. Moreover, the density
decreases to the middle of trajectory, especially for Homestake.
This is related to thicker crust in the middle of a continent.

\section{IV. \ Predictions for future experiments}

We compute the oscillation probability during a day time,
$P_{D}(E)$, according
to Eq.~(\ref{PD}). The rate of events is found using Eq. (\ref{eq:nddiff2c})
for $\nu N-$ scattering and Eq. (\ref{eq:nev3sim})
for the $\nu e-$ scattering.
The excess of night event rate was computed using
expression in (\ref{eq:nddiff3}) for the $\nu N-$ scattering and
the one in (\ref{eq:neint}) for the $\nu e-$ scattering.
These expressions correspond
to $\Delta {P}$ with neglected $I_2$, while
the phase was computed keeping the $\epsilon$ correction.

In computations we use the  Gaussian functions for  $G_{\nu N}(E^r, E)$ and
$G_{\nu e}(E^r, E)$ with certain values of the relative widths,
$\sigma_E/E$.
The nadir angle and $A_{ND}(\eta, \phi_a)$ are computed
with one minute time intervals during a year.
Then we averaged $A_{ND}(\eta, \phi_a)$ over the azimuthal angle
$\phi_a$.


We performed integration over the energies  of produced electrons
above certain thresholds. In principle, using narrow energy intervals could  improve 
the energy resolution, and consequently,  sensitivity to remote structures. 
Notice however, that with increase of neutrino energy  
the Earth matter effect increases and the resolution improves. 
Therefore due to  restricted statistics and presence of a background  the optimal 
for tomography is integration of events over energy above relatively high  threshold. 
(E.g. for DUNE we use $E^{th} = 11$ MeV.)

\begin{figure}[h]
\begin{center}
\includegraphics[width=0.45\textwidth, height=0.35\textwidth]{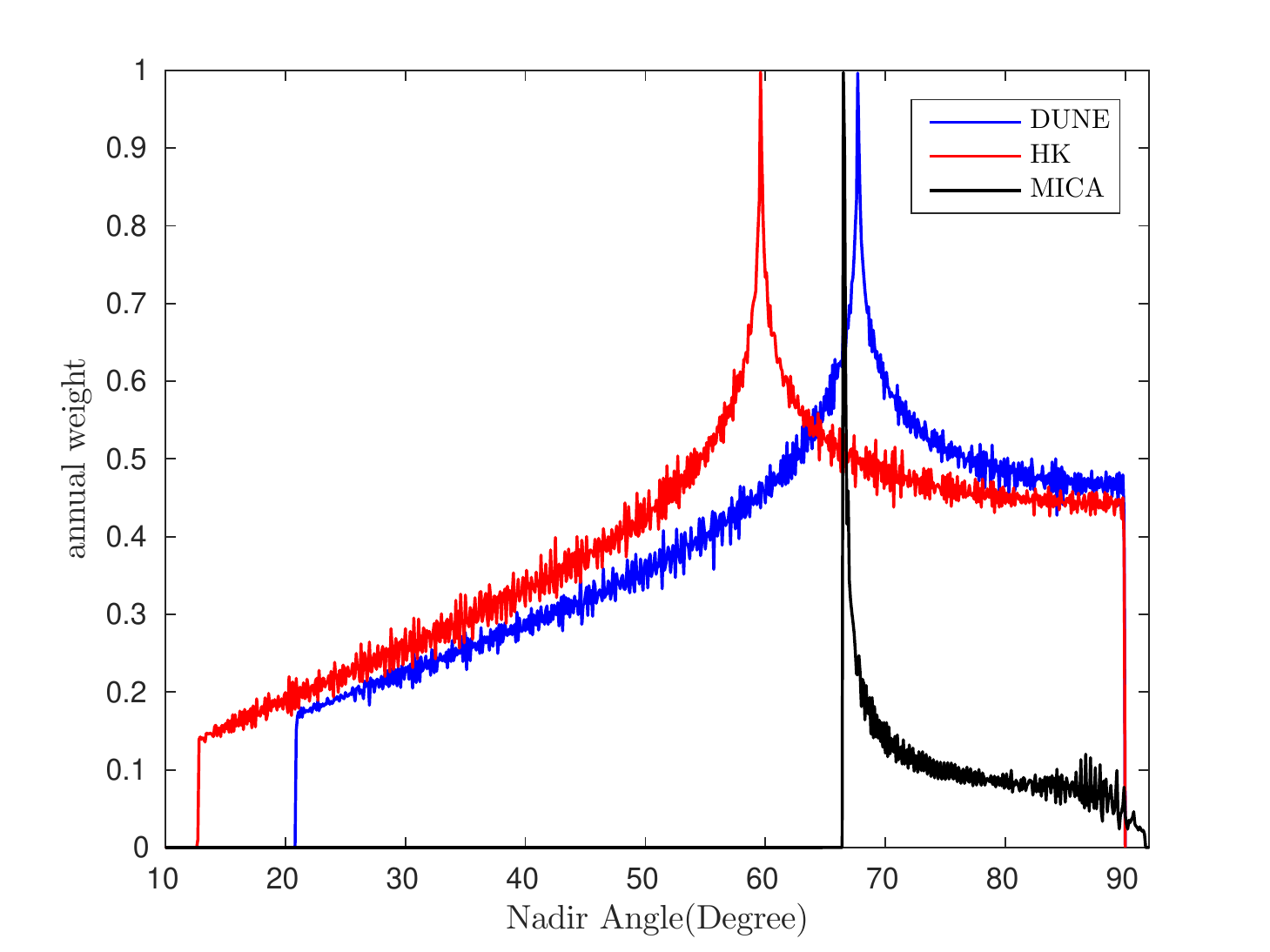}
\end{center}
\caption[...]{
Annual exposure of the detector as a function of nadir angle for
the detectors at Homestake mine, Hida Kamioka and South pole
with the time resolution of one minute and nadir
angle resolution of 0.1$^\circ$.
\label{exposure}
}
\end{figure}

We compute numerically the annual exposures for detectors at Homestake, Hida, and MICA
as functions of nadir angle
with $\Delta\eta=0.1^\circ$ (see Fig.~\ref{exposure}).
The exposure functions for Homestake is
in agreement with that in Ref.~\cite{Ioannisian:2017dkx}.
The asymmetry averaged over the year is given by integration of
${A}_{DN}$ with the exposure (weight)
function $W(\eta)$ over $\eta$:
$$
\bar{A}_{DN} = \int d \eta W (\eta) {A}_{DN} (\eta).
$$
We used exposure functions to compute the expected
experimental errors for different $\eta-$ intervals.
The value $\Delta m^2_{21} = 7.5\times10^{-5}~{\rm eV}^2$
is used unless specially indicated.

\subsection{DUNE}

DUNE is the $40$ kt liquid argon TPC which may detect solar neutrinos via
the charged current process
\begin{equation}
\nu_e + ^{40}{\rm Ar} \rightarrow ^{40}{\rm K} + e^- .
\label{eq:argon}
\end{equation}
For this process we use a generic form of cross-section
\begin{equation}
\sigma_{CC} (E) = A {p_e E_e},
\label{eq:crossect}
\end{equation}
where $A$ is a factor irrelevant for the relative excess,
$p_e$ is the momentum and $E_e = E_\nu - \Delta M$ is the energy of electron
with $\Delta M = 5.8~\rm{MeV}$ being
the reaction threshold \cite{Ioannisian:2017dkx}.
Only 9.7\% of $^8$B neutrinos have energy
$E_\nu$ $ >$ 11~\rm{MeV} but due to strong energy
dependence in (\ref{eq:crossect})
the corresponding fraction of detected events is 0.9. Therefore, we use
the threshold 11 MeV to achieve higher energy reconstruction. For resolution functions $g_\nu$ that enter $G_{\nu N}$ 
we use $\sigma_E/E_e = 0.1$. 
With this parameters  the width of the generalized resolution function  $G_{\nu N}$ turns out to be  
$\sigma_E/E = 7\%$, and consequently,
the attenuation length equals
$\lambda_{att} = 1800$ km for the average energy 12 MeV.
The nadir angle at which the length of trajectory
$L > \lambda_{att}$ is $\eta_{att} = 82^\circ$.
For $\eta < \eta_{att}$ the Earth structures on the remote part of a neutrino trajectory become invisible.

\begin{figure}[h]
\begin{center}
\includegraphics[width=0.45\textwidth, height=0.35\textwidth]{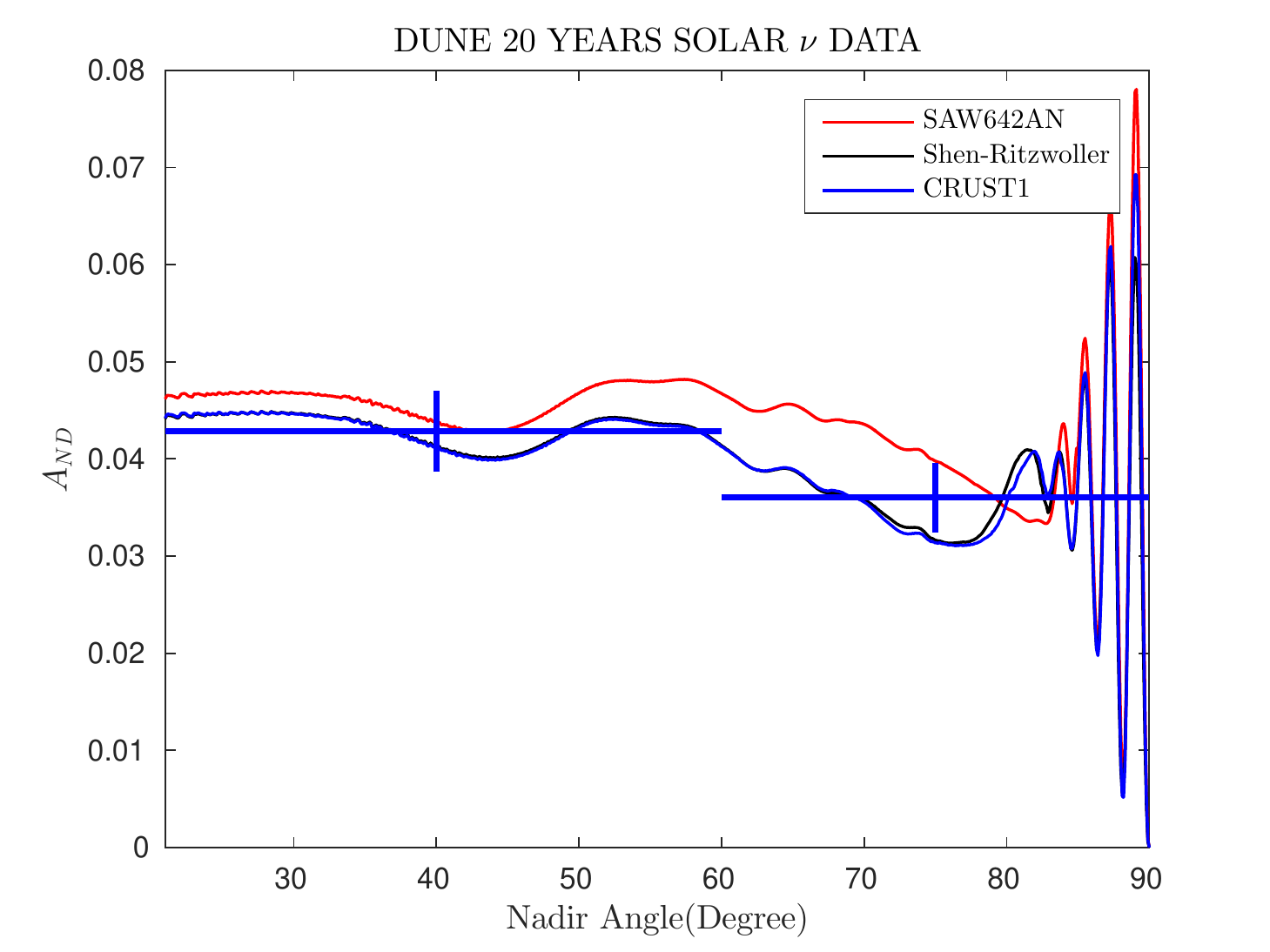}\\
\end{center}
\caption[...]{
The Night-Day asymmetry at DUNE as a function of nadir angle
for SAW642AN, Shen-Ritzwoller, and CRUST1 models.
The crosses present the expected $1\sigma$ accuracy of measurements after
twenty years of exposure for the CRUST1 model.
\label{eq:scandune}
}
\end{figure}

Results of computations of $A_{ND}(\eta)$ with
the S-R, CRUST1 and SAW642AN density profiles
are presented in Fig.~\ref{eq:scandune}.

Generic features of the $\eta$ dependence of $A_{ND}$
are the following:

(i) Oscillations in crust: Regular oscillatory pattern for $\eta > \eta_{Moho}$,
{\it i.e.} $\eta \sim 85^{\circ} - 90^{\circ}$
with decreasing depth due to averaging.
The third oscillatory peak can be affected by small
density jumps in the crust.
This quasi-regular oscillatory pattern is broken at
at $\eta_{Moho}$.

(ii) Moho interference: At $\eta < \eta_{Moho}$ neutrino trajectory crosses 
the Moho border twice leading to interference of oscillation waves from two crossings.
For some models and values of $\Delta m^2_{21}$ the destructive interference of the waves
leads to a dip at $\eta_{dip}$ (for DUNE) which depends on $\eta_{Moho}$.
This can also be interpreted as a parametric suppression of oscillations
\cite{Ioannisian:2017dkx}.

(iii) Rise of asymmetry: For $\eta < \eta_{dip}$,
the asymmetry $A_{ND}$ increases with decrease of $\eta$.
The increase is due to the fact that for small $\eta$
the section of the neutrino trajectory in the crust becomes
much smaller than the oscillation length, and so
the effective initial and final densities (averaged over the oscillation length) become
larger, being determined by the mantle density.

(iv). In the region $\eta < \eta_{dip}$ there are bump and another dip due to effect
of density jumps in the mantle at the depths 400 and 670 km.

(v) The core of the Earth $\eta_{core} = 33^\circ$ is not seen practically,
producing $\sim \epsilon^2$ effect at $\eta < \eta_{core}$.

We find that about 27000 $\nu_e$ events (\ref{eq:argon}) can be detected
annually with $E_\nu > 11~{\rm MeV}$ in the 40 kt fiducial volume according
to the CRUST1 model. Our results are comparable to Ref.~\cite{Ioannisian:2017dkx,Acciarri:2015uup,Capozzi:2018dat}. The crosses show the expected errors of $A_{ND}(\eta)$
after twenty years of data taking.
Statistical errors (computed using the exposure function)
are taken into account only and no background was considered.
As follows from Fig.~\ref{eq:scandune}, the largest
difference between SAW642AN and S-R models as well as SAW642AN and CRUST1,
is in the interval $\eta = 60^{\circ} - 77^{\circ}$ and
it originates mainly from different depths of Moho.
The difference equals $\Delta A_{ND}(\eta) \sim 0.008$ (15$\%$) which is
about 2$\sigma$ C.L., after 20 years of data taking.
The difference between CRUST1 and S-R models is practically negligible.
Averaging of $A_{ND}(\eta)$ over $\eta$ leads to $\bar{A}_{ND} = $ 0.040, 0.040 and 0.043,
for CRUST1, S-R, and SAW642AN models, respectively,
and precision of measurement of $\bar{A}_{ND}$ will be 0.002. 


New models of the Earth density profile have no spherical symmetry especially in the 
crust and upper mantle therefore 
inclusion of the azimuth angle ($\phi_a$) dependence of the density profiles in consideration should 
improve sensitivity  to specific models. 
To illustrate this we divided whole the range of $\phi_a$ in to two  bins:  
one bin is  to the west and another one to the east from a detector 
in addition to two 
nadir angle bins shown  in Fig.~\ref{eq:scandune}.  Assuming  the S-R (or CRUST1) model as the true model, 
we find that SAW642AN will be disfavored at more than 2$\sigma$ level, 
after 20 years of data taking. Integration  over the azimuth angle 
reduces the sensitivity down to  1.6$\sigma$. 
Due to low statistics in each bin introduction of more than 
two $\phi_a$  bins will not lead to further improvement of the sensitivity.

The dependence of $A_{ND}$ on $\eta$ in DUNE experiment computed with SAW642AN model
(red line Fig.~\ref{eq:scandune}) is similar to that in \cite{Ioannisian:2017dkx}
for the PREM model. It has a dip at $\eta_{dip} = 82^{\circ}$ and then
increase of $A_{ND}$ with decrease of $\eta$.
Another dip appears at $\eta = 44^{\circ}$. In our present computations
(SAW642AN) the dependence $A_{ND}(\eta)$ is smoother than in \cite{Ioannisian:2017dkx}
below the dip.

\subsection{THEIA}

THEIA is a proposed 100 kT water-based liquid scintillator detector
loaded with 1$\%$ $^7$Li \cite{Askins:2019oqj}. It will be placed
in Homestake. Neutrinos can be detected by the charged-current process
\begin{equation}
\nu_e + ^7{\rm Li} \rightarrow ^7{\rm Be} + e.
\end{equation}
The cross-section of this process is known with high precision
\cite{Alonso:2014fwf, Askins:2019oqj}. About 17000 events are expected annually
with $E_\nu$ $ >$ 5~\rm{MeV}. In the case of neutrino detection with $^7Li$, we assume $\sigma_E/E = 12\%$.

\begin{figure}[h]
\begin{center}
\includegraphics[width=0.45\textwidth, height=0.35\textwidth]{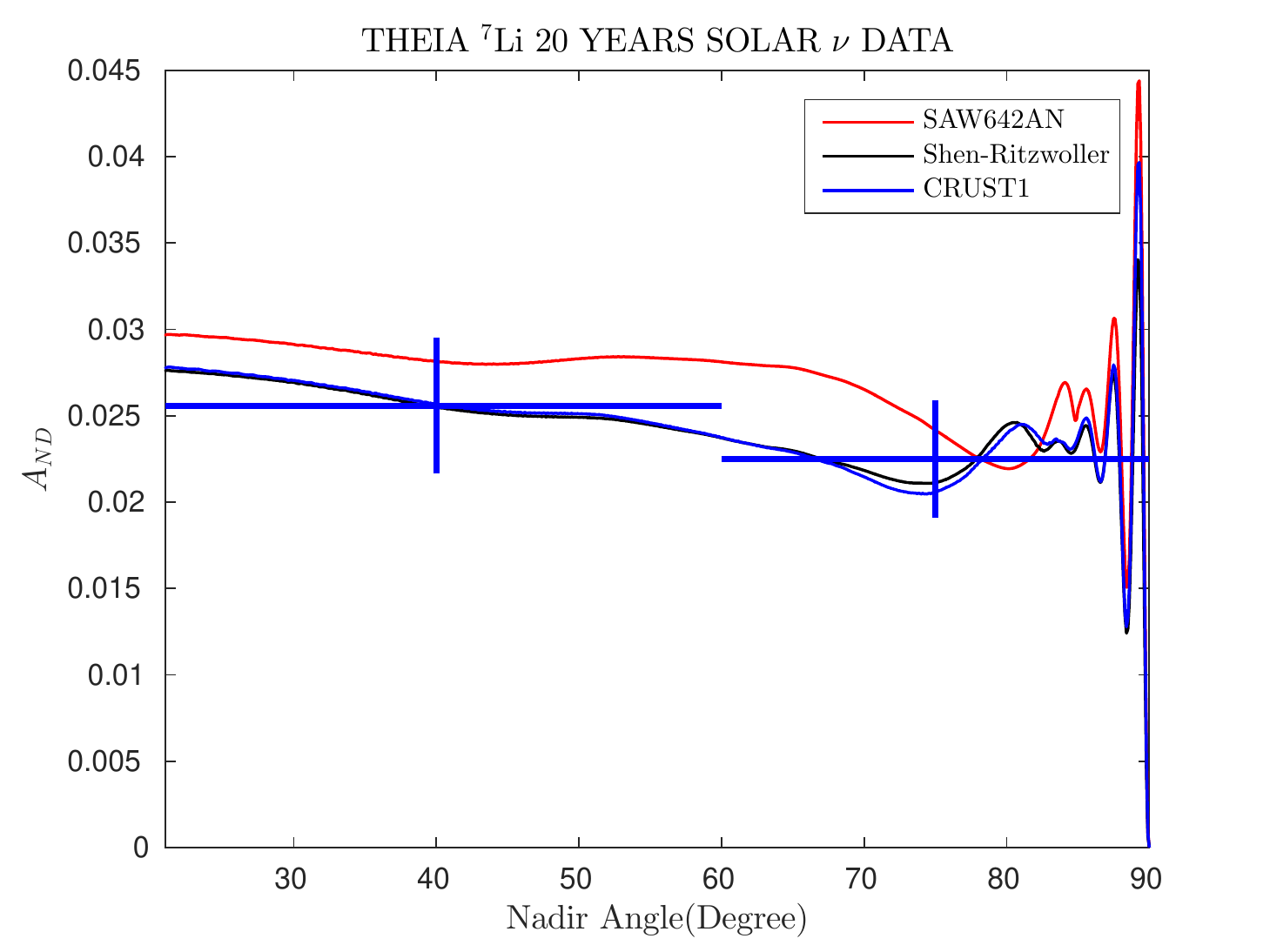}\\
\includegraphics[width=0.45\textwidth, height=0.35\textwidth]{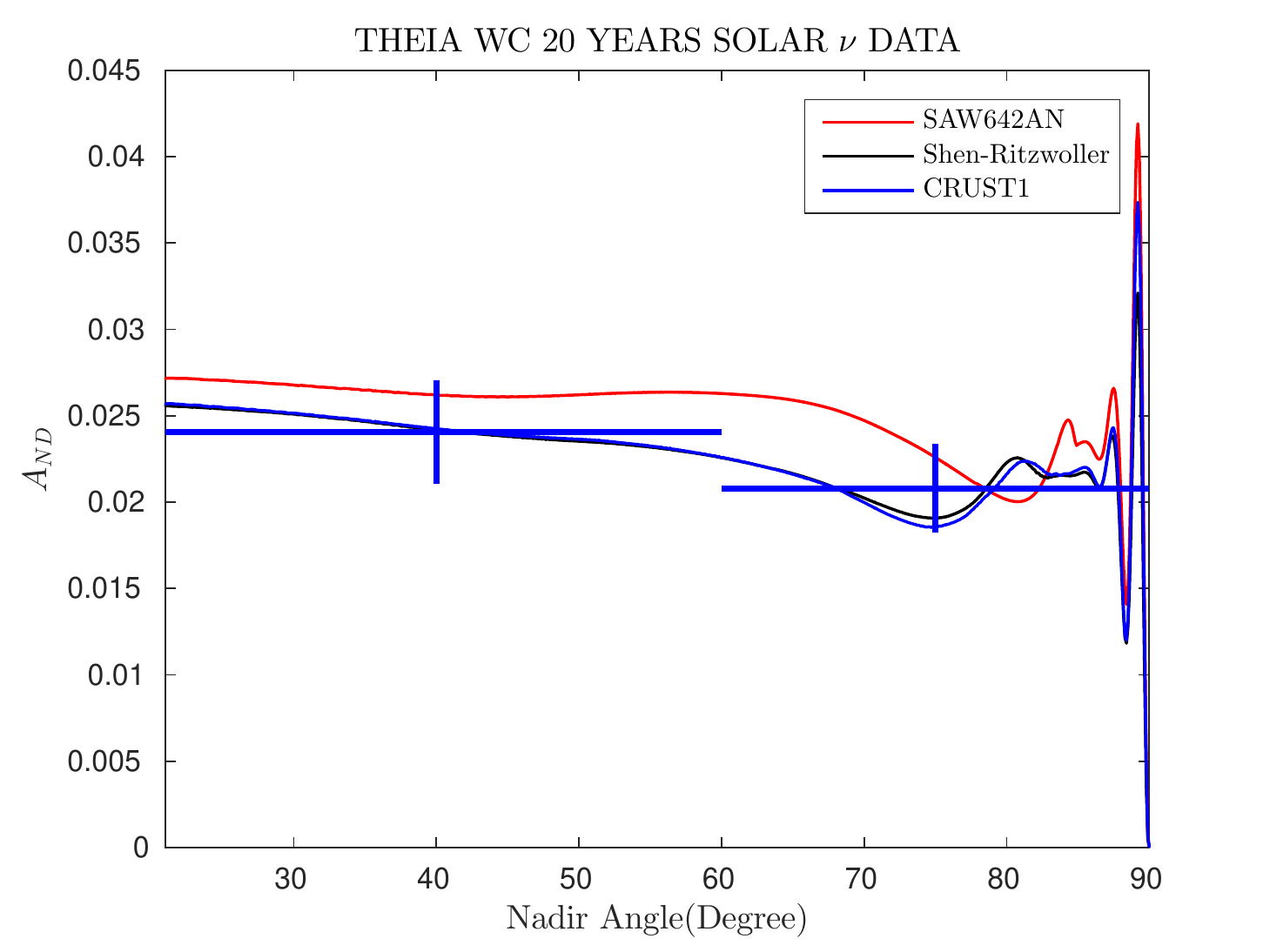}\\
\end{center}
\caption[...]{
The same as in Fig.~\ref{eq:scandune}, but for THEIA, for detection of neutrinos 
with $^7Li$ nuclei (upper panel) and elastic scattering on  electron (bottom panel).
\label{scantheia}
}
\end{figure}

Since THEIA and DUNE are in the same place the
results for $A_{ND} (\eta)$ are similar (see Fig.~\ref{scantheia}, upper panel).
The difference between $A_{ND}$ in THEIA and DUNE is due to
lower energy threshold in THEIA, which means that effective neutrino energy,
and consequently, the oscillation as well as
the attenuation lengths are smaller.
This, in turn, leads to different
interference effects and lower sensitivity to remote structures in THEIA.
The difference disappears when the same energy thresholds are taken.

For THEIA maximal difference of $A_{ND} (\eta)$ computed
with S-R and SAW642AN models
(and also between CRUST1 and SAW642AN) is about
$A_{ND} = 0.005$. 
The difference between S-R and CRUST1 profile results is much smaller.
The values of $A_{ND}$ averaged over $\eta$ with exposure taken into account in the case of $^7Li$ nuclei detection
equal to 0.024 (CRUST1), 0.024 (S-R) and 0.027 (SAW642AN).  

In THEIA neutrinos can also be detected via
the $\nu - e$ elastic scattering.  The asymmetry $A_{ND}$ as function of $\eta$  
Fig.~\ref{scantheia}, bottom panel
is similar to that for $\nu ^7{\rm Li}$ detection. 
Assuming the  energy threshold of 6.5~\rm{MeV}  and $\sigma_E/E = 0.15$, similar to HK~\cite{Hyper-Kamiokande:2016dsw},  
we find that  $A_{ND}$ equals to 0.022 (CRUST1, S-R) and 0.025 (SAW642AN), i.e. slightly smaller 
than for $\nu ^7{\rm Li}$. 
Separately, $^7{\rm Li}-$  and  $\nu  e-$ detection  can discriminate 
Shen-Ritzwoller (or CRUST1) from  SAW642AN at about  $1.6\sigma$ C.L.. 
Combining the  $^7Li$  and $\nu e$ results 
one can disfavor  SAW642AN at  more than 2$\sigma$ C.L.. 
Further combining THEIA and DUNE results, SAW642AN will be disfavored at 2.3$\sigma$ level after 20 years of data taking.

The discrimination between the S-R and CRUST1 models can be improved if
for each nadir angle $\eta$ the range of azimuthal angle $\phi_a$
is divided into two parts: in the
first part $\bar{\rho}_{SR}>\bar{\rho}_{CRUST1}$, and in the second one
$\bar{\rho}_{SR}<\bar{\rho}_{CRUST1}$. Then calculating
$A_{ND}$ in each of these parts separately and
summing up moduli of differences one can avoid averaging.

\subsection{Hyper-Kamiokande}

Hyper-Kamiokande (HK) will detect the solar neutrinos by
the $\nu - e$ elastic scattering with 6.5~\rm{MeV}
threshold \cite{Hyper-Kamiokande:2016dsw}. We take
$\sigma_E/E = 15\%$ as a tentative value.
This gives the attenuation length
$\lambda_{att} = 700$ km for $E = 10$ MeV.

\begin{figure}[h]
\begin{center}
\includegraphics[width=0.45\textwidth, height=0.35\textwidth]{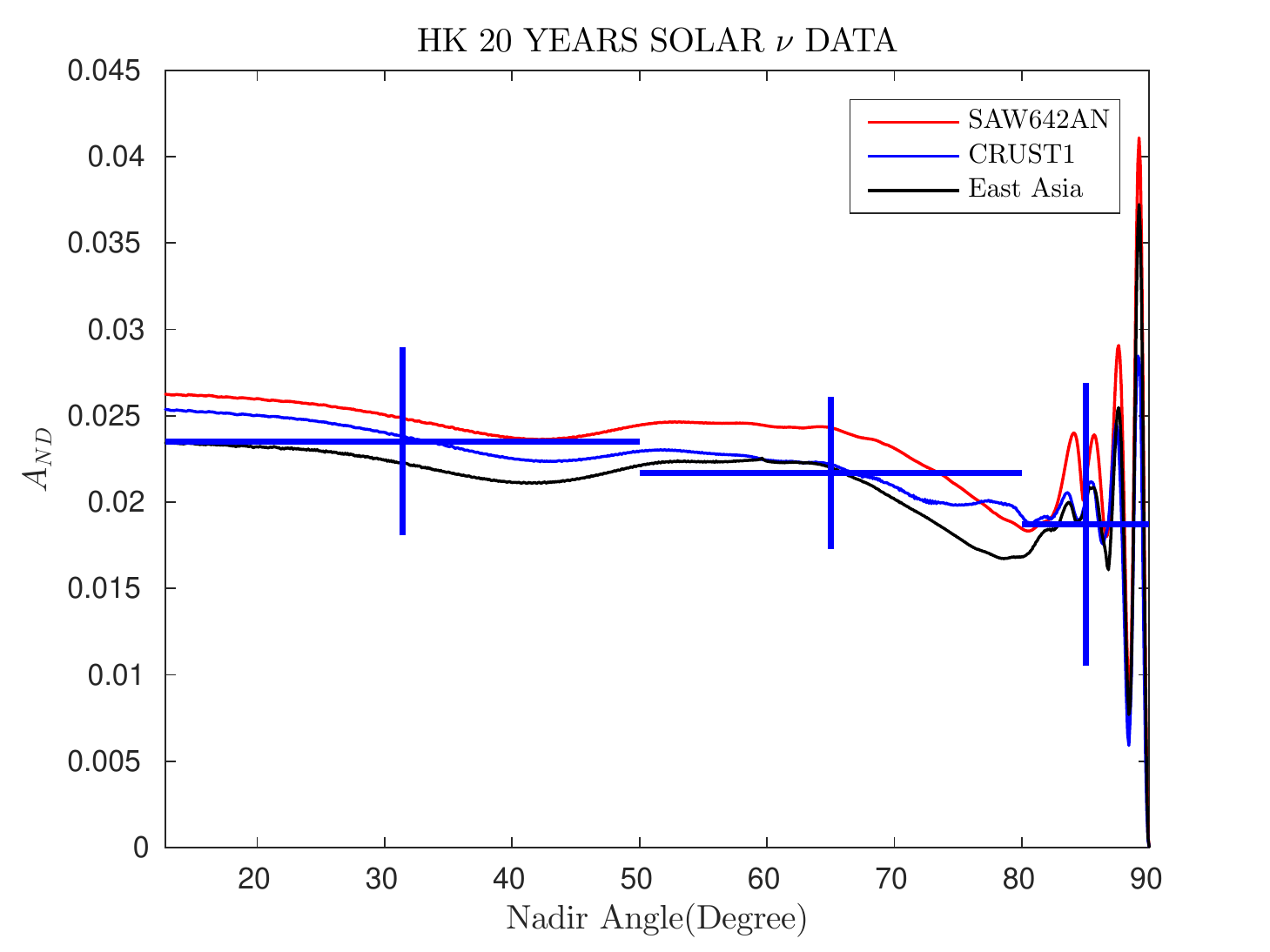}\\
\end{center}
\caption[...]{
The Day-night asymmetry at Hyper-Kamiokande as function of
the nadir angle for CRUST1, FWEA18 and SAW642AN profiles.
The crosses present expected accuracy of measurements after
twenty years of exposure taking CRUST1 as the true model.
\label{scanhk}
}
\end{figure}

In Fig.~\ref{scanhk}, we show the excess of night events
computed with FWEA18, SAW642AN and CRUST1
density profiles. For $d_{Moho} = 33$ km (FWEA18) the nadir angle
$\eta_{Moho} = 84.15^{\circ}$, and the length of the
trajectory $L = 1300$ km, so, remote half of this trajectory will not
contribute to the oscillation effect. The dip appears at
$\eta_{dip} = 78^\circ$ which is intermediate between CRUST1 and SAW64AN.

According to Fig. \ref{scanhk} maximal difference of $A_{ND}$ in HK computed
with FWEA18 and SAW642AN: $\Delta A_{ND} = 0.003$,
appears in the wide range of nadir angles: $\eta = 10^\circ - 80^\circ$.
For SAW642AN model the $\eta$ dependence in HK is similar to that
in THEIA detector. CRUST1 and FWEA18 have the biggest difference
$\Delta A_{ND} = 0.004$ in narrow range $\eta = 75^\circ - 80^\circ$.
Notice that CRUST1 does not produce the dip which is a
model-dependent feature.
The expected averaged asymmetry $A_{ND}$ in HK
equals 0.020 (FWEA18), 0.022 (CRUST1) and 0.024 (SAW642AN).
Precision of measurements of $\bar{A}_{ND}$ will be 0.002 after 20 years of 
exposure with fiducial volume 225 kton. We have considered three bins for nadir angle 
as demonstrated in Fig.~\ref{scanhk}. HK will distinguish between East Asia model  and SAW642AN, 
with 1.5$\sigma$, while CRUST1 model is recognizable from East Asia and SAW642 with 0.7$\sigma$ 
and 1.2$\sigma$ respectively after 20 years of data taking.

The absolute value of asymmetry is substantially smaller
than that for DUNE for two reasons: damping due to contribution from NC
scattering, which is 0.76, and difference of averaged energies
$E_{HK}/E_{DUNE} = 0.75$.

\subsection{MICA}

The Megaton scale Ice Cherenkov Array (MICA) is a
proposed detector at Amundsen-Scott South Pole station \cite{Boser:2013oaa}
in the same place as ICECUBE.
The latitude and longitude of MICA are 89.99$^\circ$
south and 63.45$^\circ$ west correspondingly.
Crustal structures under Antarctica are not well known
due to a lack of seismic data \cite{crust2ant},
and therefore it is interesting to explore potential
of a solar neutrino detector to determine this structure.

The detection is based on the $\nu -e$ elastic scattering.
In our calculations, we took the characteristics of
MICA from Ref.~\cite{Boser:2013oaa}: 10 Mton fiducial mass and 10~\rm{MeV}
energy threshold for the kinetic energy
of the recoil electron. With these parameters,
we find that about 5$\times 10^5$ solar
$\nu e -$ scattering events are expected per year.
For the energy resolution we use $\sigma_E/E = 15\%$.
We consider the MICA detector at a depth of 2.25 km below the icecap (as the Deep Core).
The height of icecap at the location of MICA is 2.7 km above the sea level.

\begin{figure}[h]
\begin{center}
\includegraphics[width=0.45\textwidth, height=0.35\textwidth]{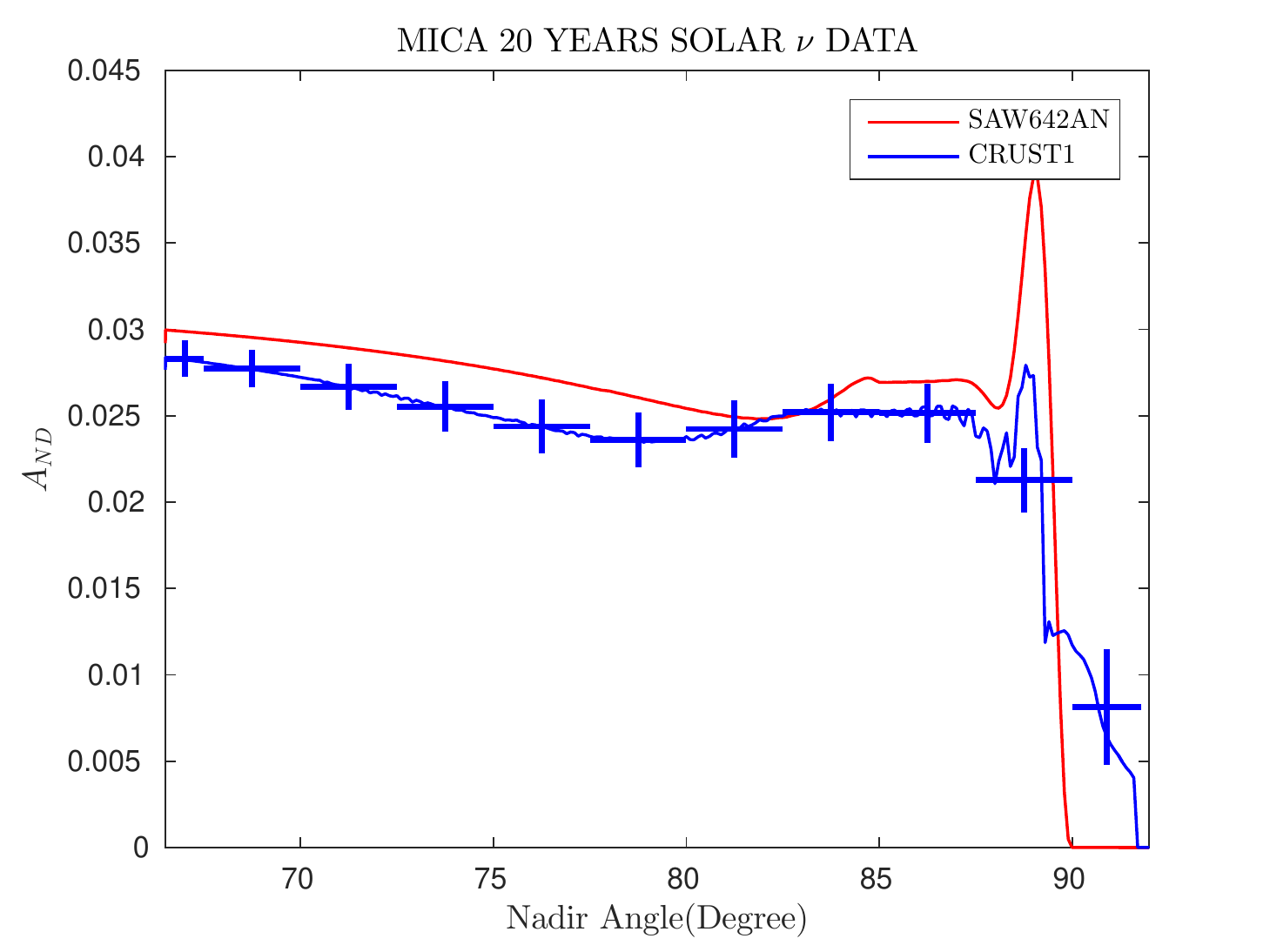}\\
\includegraphics[width=0.45\textwidth, height=0.35\textwidth]{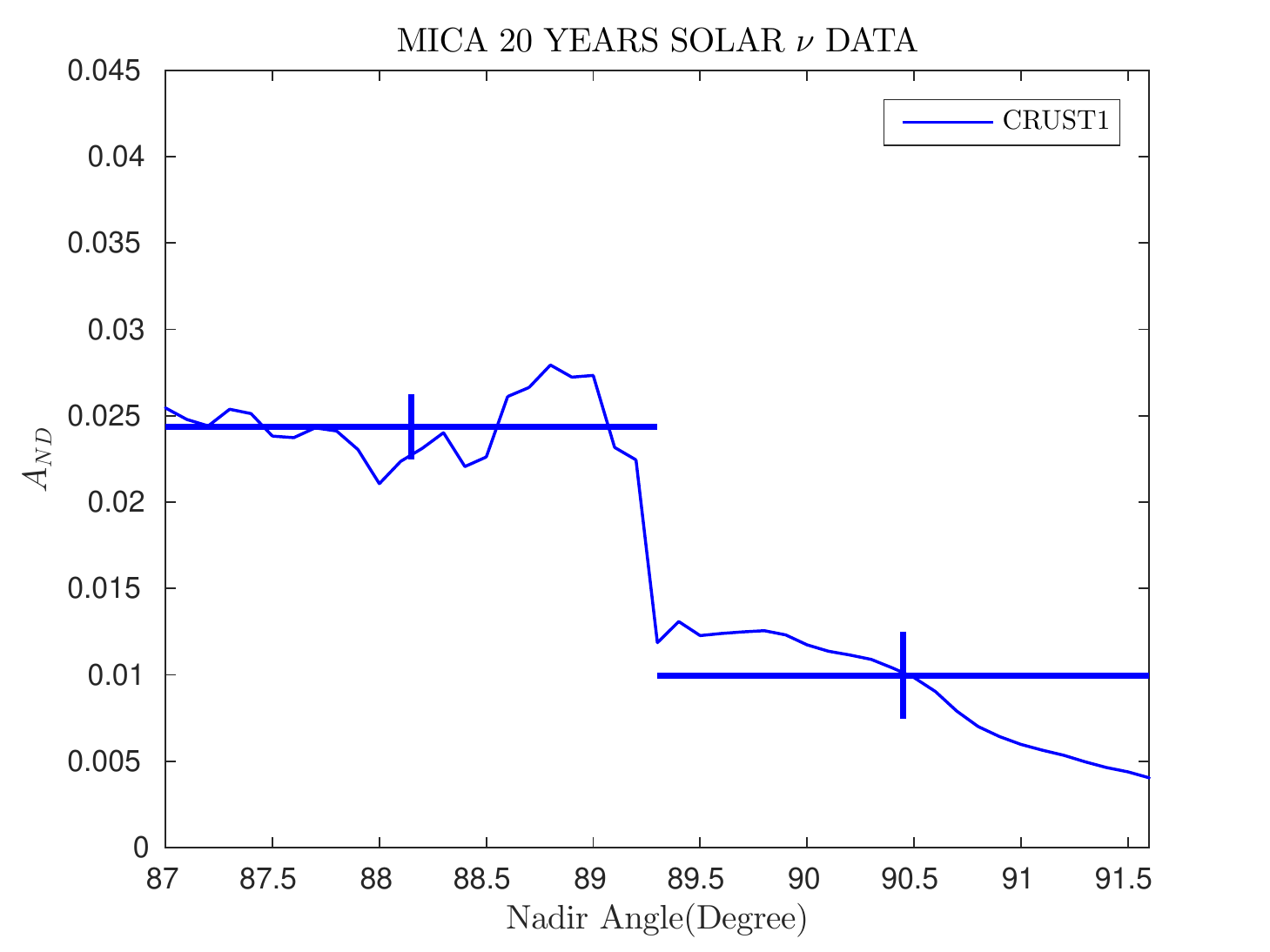}\\
\end{center}
\caption[...]{
The Night-Day asymmetry at MICA as function of $\eta$ for the SAW642AN and CRUST1 models.
The crosses present
the expected accuracy after
twenty years of exposure and taking the CRUST1 model.
Bottom panel shows zoomed part of the upper plot for nadir angles
larger than 87$^\circ$.
For $\eta > 89.3^\circ$, neutrinos cross ice only.
\label{scanMICA}
}
\end{figure}

The smallest nadir angle for MICA is 66.5$^\circ$.
About 35$\%$ of the neutrinos have the nadir angle
in the interval $66.5^\circ - 70^\circ$. These neutrinos propagate
through the Earth with a maximal depth of 500~km.
For $\eta = 75^\circ$ (where the largest difference of $A_{ND}$ from CRUST1 and SAW642AN
is expected) neutrinos propagate with a maximal
depth of 200~km. Neutrinos reach this angle on May 4
for the first time in a year. According to CRUST1 for $\eta = 75^\circ$, the depth of Moho is 35~km,
with the density jump from 2.9 to 3.4 g/cm$^3$.

In Fig.~\ref{scanMICA} we show $A_{ND}(\eta)$ computed with
CRUST1 and SAW642AN models.
CRUST1 allows taking into account the Earth density above the sea-level.
Since there is no data available for SAW642AN,
for this region, we take zero density
above the sea-level.
After 20 years of data taking MICA will collect $10^7$ solar neutrino events,
and it will be sensitive to the ice-soil border.
The average value $\bar{A}_{ND} = 0.026$ in CRUST1 model can be measured with precision 0.00045.
At $\eta > 89.3^\circ$ neutrinos pass through the ice only,
while for smaller $\eta$ they cross the ice-Earth borderline.
The SAW642AN model can be excluded
with more than 4$\sigma$, assuming that CRUST1 is true model. 

This can be further improved considering the  azimuth angle dependence of the density profile.  
For illustration in addition to 10 nadir angle bins of Fig.~\ref{scanMICA}  we introduced two equal $\phi_a$ bins:  
one to the East and another to the  West from  the detector. Analysis with 20 bins   
allows to exclude  SAW642AN at  more than 5$\sigma$.

Small ripples in $A_{ND}$ dependence on $\eta$ that appear
in the CRUST model (the blue curve
in Fig.~\ref{scanMICA}) are real.
In this model, the surface of the Earth is not spherically symmetric
and the density of the Earth above the sea-level
is given. Therefore neutrinos enter the Earth at different height from sea-level,
which leads to ripples due to change of the baseline with $\eta$.
Such ripples are far from being detected experimentally.
The ripples of $A_{ND}$ are absent in the
SAW642AN model (the red curve).

Notice that instead of the day, the cycle signal will be measured in MICA during
the year. That requires long term stability of the detector.

\subsection{Dependence on $\Delta m_{21}^2$; PREM model results}

\begin{figure}[h]
\begin{center}
\includegraphics[width=0.45\textwidth, height=0.35\textwidth]{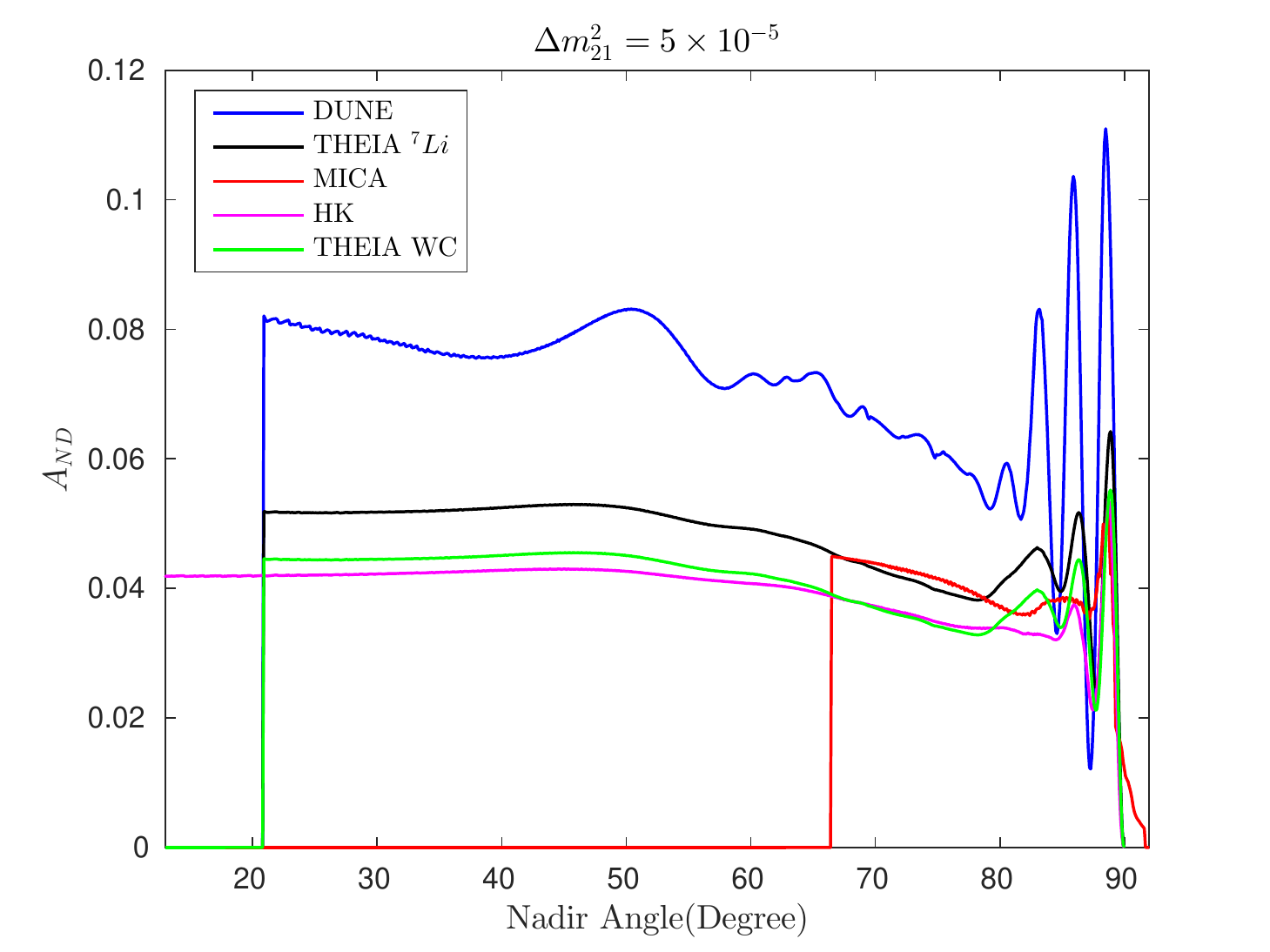}
\end{center}
\caption[...]{
$A_{ND}$ for $\Delta m^2_{21} =5\times 10^{-5}$~eV$^2$.
The S-R model was used
for DUNE and THEIA while the CRUST1 model -- for HK and MICA.
\label{dm575}
}
\end{figure}

There is a significant difference in values of $\Delta m^2_{21}$ determined by KAMLAND and from
global fit of the solar neutrino data.
In this connection we performed computations of $A_{ND}(\eta)$ using the ``solar'' value
$\Delta m^2_{21} =5\times 10^{-5}$ eV$^2$ (Fig.~\ref{dm575}).
The changes are twofold: the overall asymmetry increases
as $1/\Delta m^2_{21}$, {\it i.e.} becomes
1.5 times larger than before. The oscillation and
attenuation lengths increase by the same factor 1.5.
This, in turn, leads to (i) some change of the interference picture, (ii)
enhancement of sensitivity to remote structures and bigger densities. As a result, at small $\eta$ enhancement factor of the asymmetry is bigger than 1.5.

Let us compare results computed for DUNE with the S-R model for two different
$\Delta m^2_{21}$ (blue line in Fig. \ref{dm575} and black line in Fig. \ref{eq:scandune}).
As expected, for large $\eta$ the amplitude of oscillations of $A_{ND}$
and its average value is 1.6 times larger than those for large $\Delta m^2_{21}$.
The dip at $77^\circ$ disappears. The peak at $50^{\circ}$ is higher by a factor
$1.8$. For deeper trajectories (smaller $\eta$) the enhancement factor is $1.80 - 1.85$.
The reason for this additional increase in the
asymmetry above factor 1.5 is that due to larger oscillation length
for deep trajectories the effective initial and final densities
(averaged over the oscillation length) become larger.
For HK and CRUST1 model the results of
$\Delta m^2_{21}$ change are similar: For shallow trajectories
the asymmetry increases by factor 1.5, while
for deep trajectories (small $\eta$) -- by factor 2.

\begin{figure}[h]
\begin{center}
\includegraphics[width=0.45\textwidth, height=0.35\textwidth]{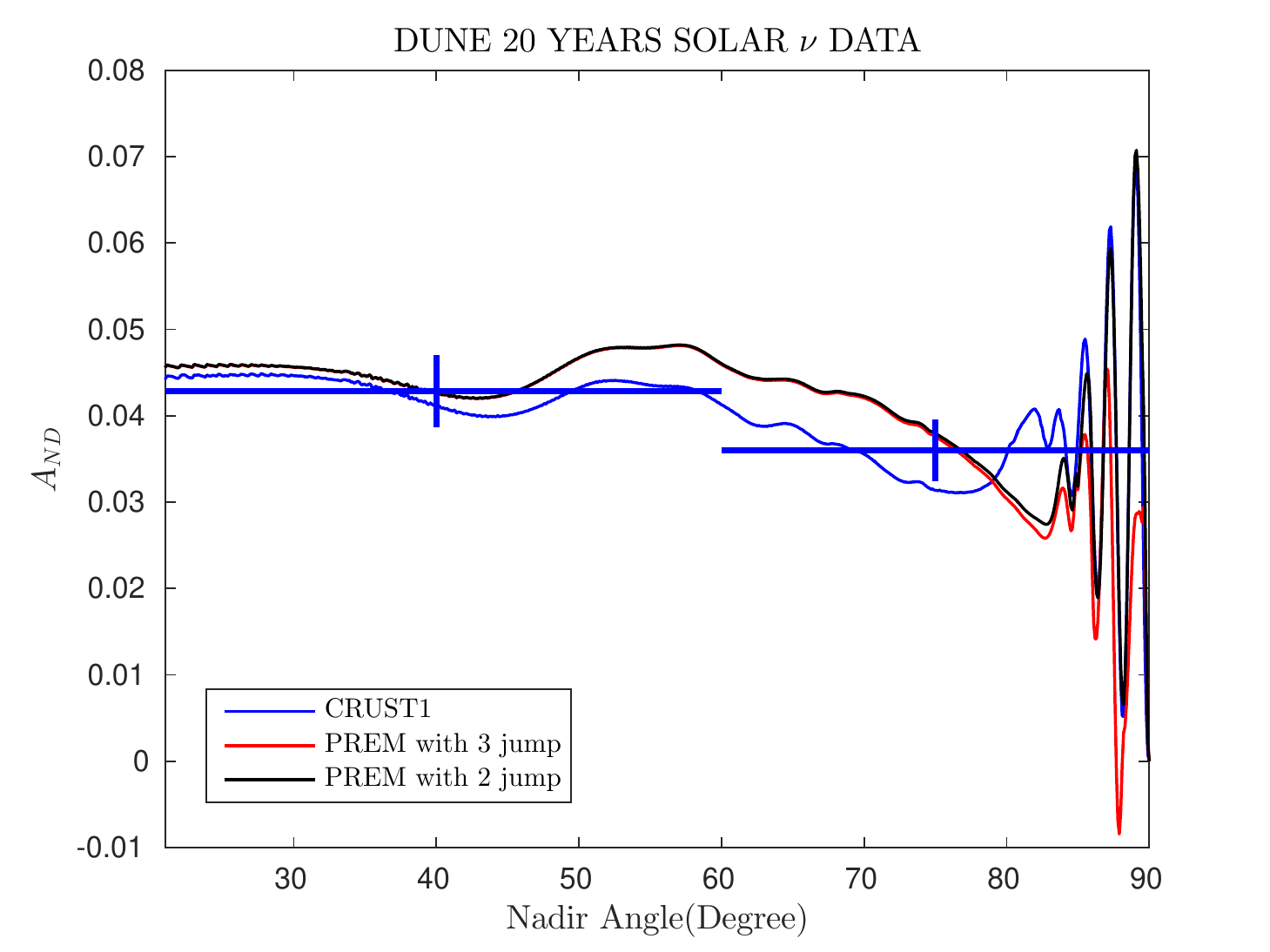}\\
\includegraphics[width=0.45\textwidth, height=0.35\textwidth]{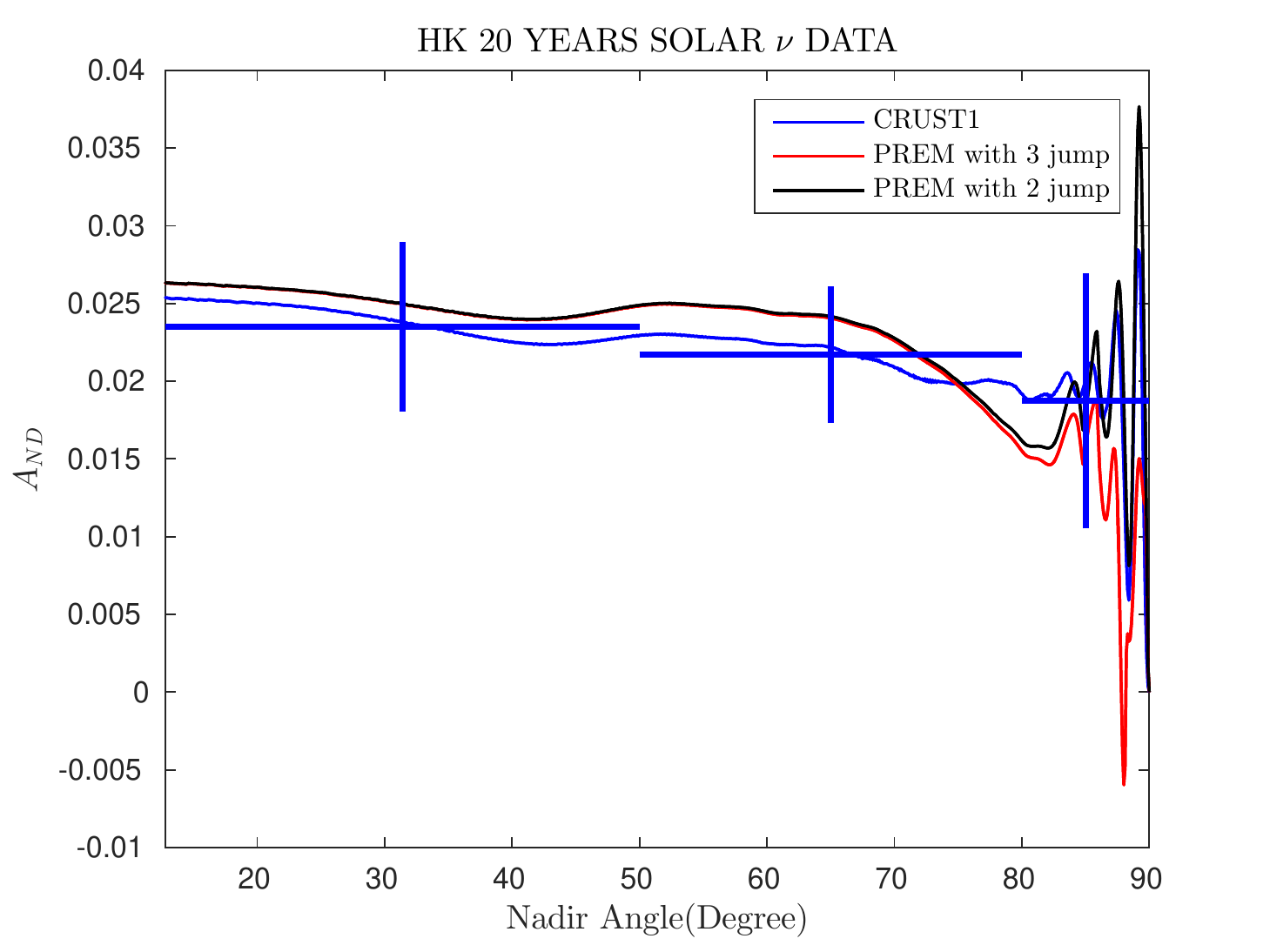}\\
\end{center}
\caption[...]{
Comparison of the $\Delta A_{ND}$ dependences on $\eta$ computed
using the PREM model with two (black line) and three (red) layers in the crust
with $\Delta A_{ND}$ dependence for the CRUST model (blue).
{\it Upper panel:} for DUNE; {\it bottom panel:} for HyperKamiokande.
\label{dANDprem}
}
\end{figure}

Notice that using new models of the Earth does not relax the tension
between the
solar and KamLAND values of $\Delta m^2_{21}$. The tension is partially
related to the fact that Super-Kamiokande found larger D-N asymmetry
than it is expected for $\Delta m^2_{21}$ given by KamLAND. In fact,
the situation with SK is similar to that for HK. According to Fig.
\ref{dANDprem}
the averaged $A_{ND}$ computed with CRUST1 model is about $5\%$ smaller
than that with the PREM model. The FWEA18 (East Asia) model gives even smaller $A_{ND}$.

\begin{figure}[h]
\begin{center}
\includegraphics[width=0.45\textwidth, height=0.35\textwidth]{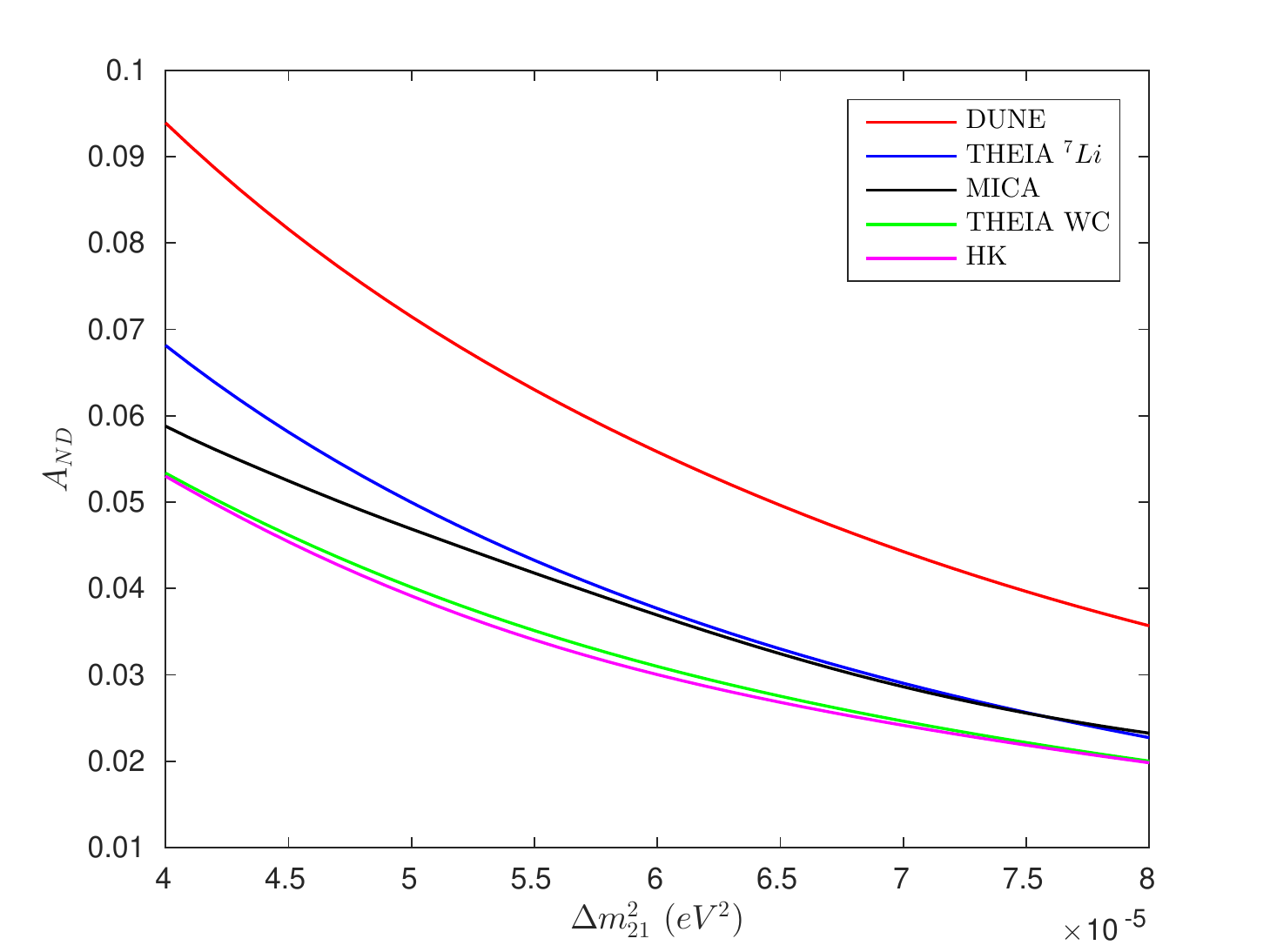}
\end{center}
\caption[...]{
The averaged over energy $A_{ND}$ as function of $\Delta m^2_{21}$
for DUNE and THEIA using the S-R model, and for HK and MICA with CRUST1 model.
\label{ANDdm}
}
\end{figure}

Most of the previous computations were performed
with PREM model which has two layers in the crust
(0 - 15) km and (15 - 24.4) km and density jumps from 2.6 to 2.9 g/cm$^3$ at 15 km,
and 2.9 to 3.38 g/cm$^3$ at 24.4 km (Moho).
The 3 km layer of water is neglected.
In Fig. \ref{dANDprem} (upper panel) we compare results of PREM (black line)
and CRUST1 (blue line) models for DUNE. The difference is mainly related to the depths of Moho:
$\eta_{Moho} = 48$ km for CRUST1, which is two times larger than
in PREM. Correspondingly, in the CRUST1 model, the dip of $A_{ND}$ is shifted to
smaller $\eta$ and for $\eta < \eta_{dip}$
the asymmetry is smaller. The latter is due to smaller effective density
(averaged over the oscillation length) near the detector in CRUST1.

The PREM result is similar to that in \cite{Ioannisian:2017dkx}.
Less profound oscillatory modulations than in \cite{Ioannisian:2017dkx}
are related to different treatment of the energy resolution.
As we mentioned before, the PREM model result is close to that
of SAW642AN model which has a similar depth of Moho.

For comparison in Fig.~\ref{dANDprem} we show also result for PREM model with
outer water layer. That would correspond to a detector near
the ocean cost. Large difference appears for $\eta > 88^\circ$
{\it i.e.} for trajectories in water: the depth of oscillations
and average $\bar{A}_{ND}$ are smaller
since they correspond to small water density 1.02 g/cm$^3$.

Similar situation is for HK Fig.~\ref{dANDprem} (bottom).
According to CRUST1 the dip is absent,
$A_{ND}$ is larger in the range $\eta = 75^\circ - 85^\circ$,
while at $\eta < 75^\circ$ the asymmetry is 10\% smaller (by 0.002) than for PREM.

The results show that usage of PREM model causes up-to 10$\%$
relative systematic error in $A_{ND}$.

Another approach to the oscillation tomography is to use
the energy spectrum distortion for fixed direction $\eta$.
Inverse problem of reconstruction of the density profile from
the energy distortion was considered in \cite{Akhmedov:2005yt}. In particular, effects of deviation from spherical
symmetry were discussed using a toy model.

\section{V. Conclusion}

1. We performed detailed study of the Earth matter effects on solar
neutrinos using recent 3D models of the Earth.
Interesting and non-trivial oscillation physics is realized which is related to complicated density
profiles along neutrino trajectories.
The Day-Night asymmetry as a function of the nadir angle has been computed for
future experiments DUNE, THEIA and HyperKamiokande, as well as for possible
next-after-next generation experiment MICA.
This allows us to assess feasibility of tomography of the Earth with solar
neutrinos.

2. We estimated corrections to $A_{ND}$ of the order $\sim \epsilon^2$.
Corrections $\sim \epsilon^2$ from $I_2$
can be neglected due to additional small
coefficient, while the $\epsilon$ correction
to the oscillation phase can be relevant.

3. We further elaborated on the attenuation effect.
The night excess of events and $A_{ND}(\eta)$ are expressed
in terms of the matter potential and the generalized energy
resolution function which, in turn, determines the attenuation factor.
This form is the most appropriate for tomography.
We have found that inclusion of energy dependence of the boron neutrino flux
and cross-section into resolution function improves the resolution,
and therefore sensitivity to remote structures.
It is the generalized resolution function that determines sensitivity
of oscillation results to the density profile.

Further improvement of the sensitivity can be achieved imposing
high enough energy threshold for detected electrons. The gain is twofold:
(i) The Earth matter effect increases as $E$; (ii) the attenuation becomes weaker.
At the same time loss of statistics is rather moderate.

4. Using recently elaborated 3D models of the Earth we reconstructed
the density, and consequently, potential
profiles along neutrino trajectories characterized by coordinates of a detector,
nadir and azimuthal angles.
The key feature of the models is the absence of spherical symmetry.
Averaging over $\phi_a$ leads to dumping of oscillatory modulations.

The key feature of profiles that determines the $A_{ND}(\eta)$
is the depth of Moho (border between crust and mantle). The depth
differs substantially in different models, and furthermore, the border substantially
deviates from spherical form.

5. Difference of results for different models of the Earth
at DUNE and THEIA at Homestake is about $10\%$. After 20 years of DUNE exposure
that would correspond to
$2\sigma$ C.L.. So, the models cannot be discriminated.
Similar conclusion is valid for HK.

6. MICA will be sensitive to the ice-soil border.
It can discriminate between the CRUST1 and SAW642AN models at $5\sigma$ C.L.
after 20 years of data taking.

7. With decrease of $\Delta m^2_{21}$ the overall excess increases as $1/\Delta m^2_{21}$.
Also $\eta$ dependence changes which is related to increase of the oscillation length
and therefore decrease of the oscillation phase: for deep trajectories the enhancement
with decrease of $\Delta m^2_{21}$ is stronger than $1/\Delta m^2_{21}$.

8. The difference of results obtained for Homestake with
S-R and CRUST1 from those of
PREM model, which was used in most of the previous studies,
is that the dip in the nadir angle distribution does not
appear and for deep trajectories the asymmetry is $10\%$ lower.

In conclusion, future experiment DUNE, THEIA, HK will certainly establish
the integrated Earth matter effect
with high significance. They may observe some generic features of the $\eta$
dependence such as dip and slow increase of the excess with decrease of
$\eta$. However, they will not be able to discriminate between
recent models. For this megaton scale experiments like MICA are needed.

\subsection*{Acknowledgments}
We would like to thank the anonymous referee for her/his useful comments.
P.B. would like to thank M. Rajaee, M. Bahraminasr and M. Maltoni for useful discussions. P.B. received funding from the European Union’s Horizon 2020 research and innovation programme under the Marie Sk\l{}odowska-Curie Grant Agreement No.~674896 and No.~690575.
P.B. is supported by Iran Science Elites Federation Grant No. 11131. P.B. thanks MPIK and IFT for their kind hospitality and support.

\section*{Appendix A. \ Neutrino Trajectory in the Earth}

The Earth can be considered as a sphere with a very small
compared to the Earth radius deviations from the sphere.
So, the distance of a given point at the surface from the centre
of the Earth equals $r_E(\theta,\phi) = 6371~km+H(\theta,\phi)$,
where $H(\theta,\phi)$ is the height from the sea-level of the location.
Here, $\theta$ and $\phi$
are the latitude and longitude of the point respectively.
Let us introduce coordinates $x$, $y$ in the plane
perpendicular to the axis of rotation of the Earth and $z$ being along the axis.
The axis is tilted by about $\alpha=23.4^\circ$ relative to the Earth orbital plane.
In these coordinates location of a point on the Earth surface
at a given moment of time $t$ is determined by
\begin{eqnarray}
x & = & r_E(\theta,\phi) \cos\theta\cos(\phi+\omega t),
\label{eq:place_on_earthx}
\\
y & = & r_E(\theta,\phi) \cos\theta\sin(\phi+\omega t)\cos\alpha-r_E(\theta,\phi)\sin\alpha\sin\theta,
\nonumber
\\
z & = & r_E(\theta,\phi) \sin\theta\cos\alpha+r_E(\theta,\phi)\cos\theta\sin(\phi+\omega t)\sin\alpha,
\nonumber
\end{eqnarray}
where
$\omega$ is the angular frequency of the Earth rotation.

Location (latitude and longitude) of DUNE, THEIA
(Homestake) is $44.35^\circ$ of north and $103.75^\circ$ of the west.
For H-Kamiokande (Hida) we have $36.23^\circ$ of
north and $137.19^\circ$ of the east,
and for MICA (Amundsen-Scott South Pole Station):
89.99$^\circ$ south and 63.45$^\circ$ west.

In all the cases except for MICA
we have considered the Earth surface as a perfect sphere ($H(\theta,\phi)=0$),
and the detectors located at the surface of the Earth.
In the case of MICA, we used CRUST1 model, which allow taking into account
$H(\theta,\phi)$, and the detector
is location 2.25~$km$ below the ice surface.

The coordinates of the Earth in the solar system are
\begin{equation}
X = r_a \cos(\Omega t+\Phi_0), ~~~~
Y = r_a \sin(\Omega t+\Phi_0),
\label{eq:place_on_earth}
\end{equation}
where $\Omega$ is 2$\pi/(365.256$ days), and $r_a=a(1-b\cos\Omega t)$ is the distance between
Earth and Sun. Here a=1 is the astronomical unit,
and b=0.0167 is the eccentricity of the Earth orbit.
For the starting point, t = 0, at the 23rd of September
the phase equals $\Phi_0 = -\frac{\pi}{2}$.

Let $x_D$ and $y_D$ be the coordinates of
the detector and $x$, $y$ and $z$ are the coordinates
of the point at which neutrino enters the Earth.
The neutrino trajectory inside the Earth is determined
by solving the following quadratic equation:
\begin{equation}
x^2 + y^2 = r_D^2, ~~~~y = m (x-x_D)+y_D,
\label{eq:location}
\end{equation}
where $m \equiv Y/X$ and $r_D^2 = x_D^2+y_D^2$.
Taking into account tilt $\alpha$,
the latitude and longitude of the entering point to the Earth
and consequently, the trajectory of the neutrino
inside the Earth as well as the nadir angle are determined.

To perform a precise calculation of the neutrino trajectory
for MICA we use the CRUST1 model.
In this case, the Earth is not a perfect sphere. Therefore
we solved the quadratic equation first with $r_E$ that includes $H_d$ the depth of the detector
from the sea-level. In this way, we
obtained the entrance point of the neutrinos into the Earth, $\theta_0$ and $\phi_0$.
Then we have solved Eq. (\ref{eq:location}) once again with $H(\theta_0,\phi_0)$.

\section*{References}


\end{document}